\documentclass[12pt,english]{article}
\usepackage{
amsmath,
amssymb,
amsthm,
amsxtra,
authblk,
babel,
braket,
booktabs,
caption,
cite,
float, 
graphbox,
graphicx,
hyperref,
dsfont,
pgfplots,
siunitx,
stackengine,
subfig,
subfloat,
tabu,
xcolor,
xparse,
}
\usepackage[utf8]{inputenc}

\hypersetup{colorlinks=true,
	linkcolor={black},
    	citecolor={black},
    	urlcolor={blue!30!black}
   	 }

\addto\extrasenglish{%
}
\addto\extrasenglish{%
}
\addto\extrasenglish{%
}

\newcommand{\eqb}{\begin{equation}}
\newcommand{\eqe}{\end{equation}}
\newcommand{\dmb}{\begin{displaymath}}
\newcommand{\dme}{\end{displaymath}}

\newcommand{\eab}{\begin{eqnarray}}
\newcommand{\eae}{\end{eqnarray}}

\newcommand{\be}{\begin{equation}}
\newcommand{\ee}{\end{equation}}

\RenewDocumentCommand\[{}{\begin{equation}}
\RenewDocumentCommand\]{}{\end{equation}}
\NewDocumentCommand\der{}{\mathrm{d}}

\NewDocumentCommand\gsim{}{\gtrsim}
\NewDocumentCommand\lsim{}{\lesssim}

\NewDocumentCommand\intkern{}{\int\kern-5pt}
\NewDocumentCommand\mat{m}{\left(\begin{matrix}#1\end{matrix}\right)}
\NewDocumentCommand\lagr{}{\mathcal{L}}
\NewDocumentCommand\abgd{}{\alpha\beta\gamma\delta}
\NewDocumentCommand\badg{}{\beta\alpha\delta\gamma}
\NewDocumentCommand\projl{}{(\mathds{1}-\gamma^5)}
\NewDocumentCommand\projr{}{(\mathds{1}+\gamma^5)}

\NewDocumentCommand\plushc{}{+\mathrm{h.c.}}

\newcommand{\shrinkage}{2.1mu}
\NewDocumentCommand\dvec{m}{\def\useanchorwidth{T}\stackon[-5.2pt]{#1}{\,\smash{\stackon[-2.25pt]{\mkern-\shrinkage\mathchar"017E}{\rotatebox{180}{$\mkern-\shrinkage\mathchar"017E$}}}}}
\stackMath

\usepgfplotslibrary{groupplots}

\title{\bf General Neutrino Interactions from an Effective Field Theory Perspective}
\author{Ingolf Bischer, Werner Rodejohann}

\date{{\it 
Max-Planck-Institut f\"ur Kernphysik, \\Postfach
103980, D-69029 Heidelberg, Germany}\\
\href{mailto:bischer@mpi-hd.mpg.de}{bischer@mpi-hd.mpg.de}, 
\href{mailto:werner.rodejohann@mpi-hd.mpg.de}{werner.rodejohann@mpi-hd.mpg.de}
\\[2ex]%
}

\begin{document}
  \maketitle
\begin{abstract}
\noindent
General Neutrino Interactions (GNI) are scalar, pseudoscalar, vector, axial vector or tensor interactions of neutrinos with fermions, and generalise the often studied neutrino Non-Standard Interactions (NSI). 
If GNI arise from heavy new physics, they should be embeddable into effective field theory operators that respect the Standard Model (SM) gauge symmetry. Therefore we consider a full basis of gauge-invariant dimension-six operators involving SM fermions and right-handed singlet neutrinos and map their Wilson coefficients onto GNI parameters. In this embedding we discuss 
correlations of and bounds on different GNI in the context of charged lepton flavour violation processes and neutrino-fermion scattering, as well as beta decay and coherent neutrino-nucleus scattering. 
We also study possible UV completions of the relevant dimension-six operators for GNI via leptoquarks that can be related to radiative neutrino masses and $B$ physics anomalies. Details on the numbers of free GNI parameters for Dirac or Majorana neutrinos and for CP violation or conservation are also provided. 

\end{abstract}
\newpage
\section{Introduction}
As neutrino physics is entering the precision era, structures beyond their Standard Model (SM) interactions can be probed. This paper will focus on new neutrino interactions of scalar, pseudoscalar, vector, axial vector and tensor type, which we will denote as General Neutrino Interactions (GNI).  
Interpreting GNI as an effect of high scale new physics leads to effective field theories (EFTs) as the suitable framework to describe them in a model-independent way.

Indeed, a subset of GNI has long been parametrised by Wilson coefficients of non-renormalisable dimension-six operators, namely the so-called Neutrino Non-Standard Interactions (NSI) \cite{Ohlsson:2012kf,Farzan:2017xzy}. Those are vector-type interactions of neutrinos with charged  fermions. Like GNI, they are described via EFT operators below the electroweak scale that respect the residual $\mathrm{SU(3)}_\mathrm{C}\times\mathrm{U(1)}_\mathrm{em}$ gauge symmetry. 
On the other hand, in collider physics different EFTs have been established, most prominently the Standard Model Effective Field Theory (SMEFT), which describes physics around and above the weak scale respecting the full $\mathrm{SU(3)}_\mathrm{C}\times\mathrm{SU(2)}_{L}\times\mathrm{U(1)}_Y$ SM gauge group. In order to compare results, one needs to match the operators of the two EFTs at the weak scale which seperates the respective domains of validity. 
This has been performed in \cite{Jenkins:2017jig} including all possible operators below the weak scale.

However, it turns out that many of the GNI cannot be produced as a low-energy limit of SMEFT operators. 
In this work, we therefore extend the SMEFT with right-handed neutrinos $N$, which we abbreviate by SMNEFT. 
This is rather well motivated extension, given that neutrinos are massive. 
We find that indeed almost all GNI can be produced from four-fermion operators in the SMNEFT.
Subsequently we investigate in which cases detectable GNI and interesting correlations between different GNI parameters are possible. We choose here coherent elastic neutrino-nucleus scattering
(CE$\nu$NS) and beta decay as an attractive example. 
At the example of charged lepton flavour violation (CLFV) searches, we also demonstrate how powerfully charged-fermion observables can constrain deviations from standard neutrino-fermion scattering when interpreted in SM(N)EFT. Finally, we show that viable UV complete scenarios exist that lead to the effective SMNEFT operators after integrating out heavy leptoquarks, some of which are candidates to explain radiative neutrino masses and $B$ physics anomalies.\\

This article is structured as follows. In \autoref{sec:EFT}, we outline the construction and parametrisation of the low-energy EFT, namely GNI, and the high-energy EFT, namely SMNEFT, before discussing the operator matching.
Phenomenological aspects of this matching are discussed in \autoref{sec:pheno}. In particular, we find that constraints on CLFV place strong bounds on some flavour-changing neutrino interactions. On the other hand, we find that interesting correlations between beta decay modifications and distortions of the recoil spectrum in CE$\nu$NS could be testable. 
The UV completions via heavy leptoquarks are discussed in \autoref{sec:leptoquarks} before summarising and concluding in \autoref{sec:conclusions}.

\section{EFT below and above the weak scale}
\label{sec:EFT}
If one is to consider effects of new physics of some high scale $\Lambda$ on processes at a lower energy scale $\mu\ll\Lambda$ in a model independent way the suitable framework is an EFT composed by operators involving the fundamental (or composite) fields relevant at the scale $\mu$ and respecting the symmetry structure of the theory at this scale. Effects of the new physics may manifest themselves in non-renormalisable operators of dimension $d\geq5$, such that one can expand
\[\label{eq:eftexpansion}
\lagr_{\mathrm{eff}} = \lagr_{\mathrm{SM}}+\sum_i\sum_{n\geq5}\frac{1}{\Lambda^{n-4}}C_i\mathcal{O}^{(n)}_i\,, 
\]
where $C_i$ denotes the dimensionless Wilson coefficient of the operator $\mathcal{O}_i$, operators of higher mass dimension are suppressed by respective factors of $\Lambda^{-n}$, and $\Lambda$ is considered as the cutoff of the EFT. Here, we will be concerned with dimension-six operators suppressed by $\Lambda^{-2}$. However, we will instead parametrise the Wilson coefficients with respect to the Fermi constant $\sqrt{8}G_F$, such that they can be directly compared to the strength of the weak interactions. 

In the following, we will specify two such EFTs by introducing GNI as a low-energy example and the effective field theory of the SM above the weak scale (SMEFT) enhanced by right-handed neutrinos as a high-energy example (SMNEFT). Subsequently, we discuss the matching between both descriptions.

\begin{table}
{\centering
\begin{tabu}spread 50pt{X[$,c]X[$,c]X[$,c]X[$,c]X[$,c]}
\toprule
& \text{Spin} & \mathrm{SU(3)}_\mathrm{C} & \mathrm{SU(2)}_L & \mathrm{U(1)}_Y\\
\midrule
l & 1/2 & 1 & 2 & -1\\
N & 1/2 & 1 & 1 & 0\\
e & 1/2 & 1 & 1 & -2\\
q & 1/2 & 3 & 2 & 1/3\\
u & 1/2 & 3 & 1 & 4/3\\
d & 1/2 & 3 & 1 & -2/3\\
\varphi & 0 & 1 & 2 & 1\\
\bottomrule
\end{tabu}
\caption{Fermionic and scalar field content employed in this work. The vacuum expectation value (vev) of $\varphi$ in unitary gauge reads
$\braket{\varphi}=\frac{1}{\sqrt{2}}\mat{0,v}^T$. 
The electric charge operator is defined by $Q=I_3+Y/2$, where $I_3$ is the third generator of weak isospin and $Y$ the hypercharge operator. 
}
\label{tab:smcontent}
}
\end{table}

\subsection{General Neutrino Interactions}
\label{sec:GNI}

In this section we consider an EFT of the SM below the weak scale. Namely, dimension-six operators are composed of the SM fields with masses below the weak scale, that is, leptons, quarks (except for top), photons and gluons which transform under the gauge group $\mathrm{SU}(3)_\mathrm{C}\times\mathrm{U}(1)_\mathrm{em}$. 
However, we confine ourselves with operators involving neutrinos and exclude photon or gluon couplings.
Our notational convention for the fermion and scalar fields is summarised in \autoref{tab:smcontent}.  Imposing Lorentz and gauge invariance under strong and electromagnetic interactions while excluding four-neutrino operators due to their very hard detectability finally leaves the GNI Lagrangians\footnote{
The reader may be more familiar with a different parametrisation of Lorentz-invariant four-fermion operators due to Lee and Yang \cite{Lee:1956qn} and is in this case referred to \autoref{sec:details} for the mapping between both parametrisations.
}
\begin{align}
\lagr^{\mathrm{NC}}&=-\frac{G_F}{\sqrt{2}}\sum_{j=1}^{10}\left(\stackrel{(\sim)\ \ \,}{\epsilon_{j,f}}\right)^{\abgd}\left(\overline{\nu}_\alpha\mathcal{O}_j\nu_\beta\right)
\left(\overline{f}_\gamma\mathcal{O}_j'f_\delta\right), \label{eq:epsilonNC}\\
\lagr^{\mathrm{CC}}&=-\frac{G_FV_{\gamma\delta}}{\sqrt{2}}\sum_{j=1}^{10}\left(\stackrel{(\sim)\ \ \ \,}{\epsilon_{j,ud}}\right)^{\abgd}\left(\overline{e}_\alpha\mathcal{O}_j\nu_\beta\right)
\left(\overline{u}_\gamma\mathcal{O}_j'd_\delta\right) \plushc
\label{eq:epsilonCC},
\end{align}
where $f=e,u,d$, denoting charged leptons, up-type and down-type quarks, Greek indices run over flavour, $V$ denotes the CKM matrix, and $\epsilon_j$, $\widetilde\epsilon_j$, $\mathcal{O}_j$, and $\mathcal{O}_j'$ are given in \autoref{tab:chiral-operators}. The entries of the $\epsilon$ flavour-space matrices are dimensionless and encode the strength of an interaction type $j$ with respect to the SM Fermi interaction. The index $j$ runs from 1 to 10 because there are five Lorentz-invariant operators for general Dirac fermions, but ten for chiral fermions, see \autoref{tab:chiral-operators}. 
\begin{table}
{\centering
\begin{tabu}spread 50pt{X[$,c]X[$,c]X[$,c]X[$,c]}
\toprule
j & \stackrel{(\sim)}{\epsilon_j} & \mathcal{O}_j & \mathcal{O}_j'\\
\midrule
1&\epsilon_L & 		\gamma_\mu \projl & \gamma^\mu \projl\\
2&\tilde\epsilon_L &  \gamma_\mu \projr & \gamma^\mu \projl\\
3&\epsilon_R &		\gamma_\mu \projl & \gamma^\mu \projr\\
4&\tilde\epsilon_R &  \gamma_\mu \projr & \gamma^\mu \projr\\
5&\epsilon_S &		\projl & \mathds{1} \\
6&\tilde\epsilon_S &  \projr & \mathds{1} \\
7&-\epsilon_P &		\projl & \gamma^5 \\
8&-\tilde\epsilon_P & \projr & \gamma^5 \\
9&\epsilon_T &		\sigma_{\mu\nu} \projl & \sigma^{\mu\nu} \projl\\
10&\tilde\epsilon_T &  \sigma_{\mu\nu} \projr & \sigma^{\mu\nu} \projr\\
\bottomrule
\end{tabu}
\caption{Coupling constants and operators appearing in generic neutral-current \eqref{eq:epsilonNC} and charged-current Lagrangians \eqref{eq:epsilonCC}.}
\label{tab:chiral-operators}
}
\end{table}
Note that we take the second fermionic bilinear of each line, \eqref{eq:epsilonNC} and \eqref{eq:epsilonCC}, in the mass basis. Although we did not write a mixing matrix factor in the neutral-current term, one should note that in contrast to the SM, the interaction parameters can be different in flavour and mass basis. In the charged-current term, we followed the convention of factoring out explicitly the CKM matrix. However, in general one has to be careful about factors of $V$ when translating the interaction parameters from mass to a particular flavour basis, as discussed in the next section.
The fact that we omitted adding the hermitian conjugate in \eqref{eq:epsilonNC} can be accounted for by demanding that
\[
\begin{split}
\epsilon_L^{\abgd}=\left(\epsilon_{L}^{\badg}\right)^*,\quad
\widetilde\epsilon_L^{\abgd}=\left(\widetilde\epsilon_{L}^{\badg}\right)^*,\quad & 
\epsilon_R^{\abgd}=\left(\epsilon_{R}^{\badg}\right)^*,\quad
\widetilde\epsilon_R^{\abgd}=\left(\widetilde\epsilon_{R}^{\badg}\right)^*,\\
\epsilon_{S}^{\abgd}=\left(\widetilde\epsilon_{S}^{\badg}\right)^*,\quad
\epsilon_{P}^{\abgd}&=-\left(\widetilde\epsilon_{P}^{\badg}\right)^*,\quad
\epsilon_{T}^{\abgd}=\left(\widetilde\epsilon_{T}^{\badg}\right)^*.
\label{eq:epsilonconstraints}
\end{split}
\]

There are a few comments in order about the Lagrangians \eqref{eq:epsilonNC} and \eqref{eq:epsilonCC}.
The well-studied NSI, see \cite{Ohlsson:2012kf,Farzan:2017xzy} for reviews, constitute a subset of the GNI, namely $\epsilon_L$ and $\epsilon_R$, i.e.\ $j=1,3$;  they involve only left-handed neutrinos. 
All the other operators involve right-handed neutrinos.
The effective description by these non-NSI operators is reasonable only for the case of sufficiently light right-handed neutrinos appearing in the interaction, namely either Dirac partners or charge conjugates of the SM neutrinos. Right-handed sterile neutrinos of sufficiently low mass (not necessarily seesaw candidates) could also appear in the interaction. In this article, we will simply assume the existence of at least one such light state and investigate the implications on GNI. For definiteness and simplicity, one may imagine a right-handed electron neutrino. 

In the case of Majorana neutrinos or additional symmetries (like charge-parity or flavour), additional constraints can be set on the neutral-current parameters, reducing the number of free parameters considerably.\footnote{
In the Majorana case, there are also lepton number violating operators such as $(\bar \nu^c e) (\bar d_R u_L)$, which, however, can not be generated by dimension-six operators \cite{Bolton:2019wta}.}
This is discussed in full detail in \autoref{sec:details}. Here we only present the number of free parameters under different assumptions in \autoref{tab:freeparameters}.
It is interesting to note that the detection of certain types of interactions which are only compatible with Dirac neutrinos, e.g. flavour-conserving tensor interactions, can rule out the Majorana neutrino scenario, while the Dirac scenario cannot be ruled out in this way \cite{Rosen:1982pj,Kayser:1981nw,Rodejohann:2017vup}.
Note that in the charged-current case \eqref{eq:epsilonCC}, relations such as in \autoref{tab:freeparameters} in general do not hold.

As a final remark, we note that another attractive property of the scalar and tensor interactions is that in their presence, the magnetic moment of the neutrino could be orders of magnitude larger than in standard scenarios, even within detector reach \cite{Xu:2019dxe}. This is a consequence of the chirality flipping nature of the interaction, such that the leading diagram involving a charged-fermion loop is no longer proportional to the neutrino mass, but to the mass of the charged fermion, which is orders of magnitude larger.

\begin{table}
{\centering
\begin{tabu}{X[c]X[c,0.5]X[c]X[c]X[c]}
\toprule
&Dirac & Majorana&CP-invariant&Majorana + CP-invariant\\
\midrule
All indices free  &
810 & 432 & 423 & 225
\\ \addlinespace[10pt]
$\gamma=\delta=\mathrm{fixed}$ &
90 & 48 & 51 & 27
\\ \addlinespace[5pt]
Flavour- \\
diagonal and $\gamma=\delta=\mathrm{fixed}$ &
30 & 18 & 21 & 12
\\ 
\bottomrule
\end{tabu}
\caption{Number of free parameters in the general neutral-current Lagrangian \eqref{eq:epsilonNC} under different assumptions, for fixed charged-fermion type $f$, either electron-type lepton, up-type quark or down-type quark. By $\gamma=\delta=\mathrm{fixed}$, we mean that we take $\gamma=\delta$ and assign a definite generation number to it. By flavour-diagonal we mean that $\epsilon_{\alpha\beta\gamma\delta}\propto\delta_{\alpha\beta}$. Details are found in \autoref{sec:details}.} 
\label{tab:freeparameters}
}
\end{table}

\subsection{SMEFT with right-handed neutrinos}
\label{sec:operators}

In the previous section, we discussed GNI as an EFT below the weak scale, considering dimension-six operators that respect the residual $\mathrm{SU}(3)_\mathrm{C}\times\mathrm{U}(1)_\mathrm{em}$ gauge symmetry. However, we know that above the weak scale the full SM gauge symmetry is restored, such that a suitable EFT should respect the SM gauge structure. Accordingly, it should consist of gauge-invariant operators that are built from fields in representations of the full SM gauge group (see \autoref{tab:smcontent}). Therefore, as symbolised in \eqref{eq:eftexpansion}, we will have all the operators present in the SM Lagrangian and, in the case that we expect new physics at a scale higher than the weak scale, effects of such new physics encoded in higher-dimensional operators ($d\geq 5$) built from the same fields. This EFT above the weak scale is the previously mentioned SMEFT pioneered in \cite{Buchmuller:1985jz,warsaw}. See e.g.\ \cite{Brivio:2017vri} for a review of SMEFT and related concepts.

Ideally, one would expect that the GNI Lagrangians \eqref{eq:epsilonNC}, \eqref{eq:epsilonCC} emerge as the low-energy limit of SMEFT.
However, as pointed out in \cite{Falkowski:2019xoe}, both the GNI ansatz (or the general EFT below the weak scale therein referred to as ``weak EFT''), and the SMEFT can be considered on an independent footing, which becomes relevant if there are new particles in the mass range of tens of \si{GeV} or if the electroweak symmetry is realised non-linearly in the new physics sector. 
In this work, however, we will assume that the low-energy theory can be matched to the high-energy EFT and discuss implications of this assumption.

A non-redundant basis of SMEFT dimension-six operators was given in \cite{warsaw} whose notation conventions we adopt here.\footnote{The only change is $\mathcal{O}_{eluq}^{(1)}$ and $\mathcal{O}_{eluq}^{(3)}$ being relabelled $\mathcal{O}_{eluq}$ and $\mathcal{O}_{eluq}'$ to avoid confusion with SU(2) triplet product operators like $\mathcal{O}_{lq}^{(3)}$ and for brevity.} 
A subset of these operators is given in the upper parts of Tables \ref{tab:4fermionSMEFT} and \ref{tab:bosonSMEFT}.
However, as the SMEFT can only produce a subset of GNI connected with left-handed neutrinos \cite{Cirigliano:2009wk, Altmannshofer:2018xyo}, we also include right-handed neutrino singlets $N$, which constitute a well-motivated extension of the SM.
The additional operators involving right-handed neutrino singlets $N$ were presented in \cite{liao}, which we partially rearrange to be more suitable with respect to neutrino interactions.\footnote{In particular, we switch from the basis elements $\{\mathcal{O}_{lNqd},\mathcal{O}_{ldqN}\}$ to $\{\mathcal{O}_{lNqd},\mathcal{O}_{lNqd}'\}$.}
We are interested in those kind of operators that involve interactions of neutrinos, either with gauge bosons and the Higgs, or among four fermions, of which at least one is charged. The four-fermion operators involving neutrinos are listed in \autoref{tab:4fermionSMEFT}, and the fermion-boson-mixing operators in \autoref{tab:bosonSMEFT}, where in both cases the SMEFT operators are listed in the upper parts, and the operators involving right-handed neutrinos are listed in the lower parts. Apart from those, there exist four-fermion operators for the interactions among four neutrinos, and the two lepton- and baryon-number violating operators
\[
\begin{split}
\mathcal{O}_{qqdN}&= C^{qqdN}_{\alpha\beta\gamma\delta}(\overline{q}^i_{\alpha}\epsilon_{ij}q^{c,j}_\beta)(\overline{d_\sigma}N^c_\delta)\,,\\
\mathcal{O}_{uddN}&= C^{uddN}_{\alpha\beta\gamma\delta}(\overline{u_\alpha}d_\beta^c)(\overline{d_\sigma}N^c_\delta)\,,
\end{split}
\]
which we are going to neglect here.
\begin{table}
{\centering
\begin{tabu}{X[$,c]X[5,$,c]X[0.5,$,c]X[5,$,c]X[$,c]X[5,$,c]}
\toprule
\multicolumn2{c}{$(\overline{L}L)(\overline{L}L)\ \mathrm{and}\ (\overline{R}R)(\overline{R}R)$} & \multicolumn2{c}{$(\overline{L}L)(\overline{R}R)$}  &
\multicolumn2{c}{$(\overline{L}R)(\overline{R}L)\ \mathrm{and}\ (\overline{L}R)(\overline{L}R)$}\\
\midrule
\mathcal{O}_{ll} & (\overline{l}_{\alpha}\gamma_\mu l_{\beta})(\overline{l}_{\gamma}\gamma^\mu l_{\delta})
& \mathcal{O}_{le} & (\overline{l}_{\alpha}\gamma_\mu l_{\beta})(\overline{e}_{\gamma}\gamma^\mu e_{\delta})
& \mathcal{O}_{elqd} & (\overline{e}_{\alpha}l_{\beta}^j)(\overline{q}_{\gamma}^jd_{\delta})
\\
\mathcal{O}_{lq}^{(1)} & (\overline{l}_{\alpha}\gamma_\mu l_{\beta})(\overline{q}_{\gamma}\gamma^\mu q_{\delta})
& \mathcal{O}_{lu} & (\overline{l}_{\alpha}\gamma_\mu l_{\beta})(\overline{u}_{\gamma}\gamma^\mu u_{\delta})
& \mathcal{O}_{eluq} & (\overline{e}_{\alpha} l_{\beta}^j)\epsilon_{jk}(\overline{u}_{\gamma} q_{\delta}^k)
\\
\mathcal{O}_{lq}^{(3)} & (\overline{l}_{\alpha}\gamma_\mu\tau^I l_{\beta})(\overline{q}_{\gamma}\gamma^\mu\tau^I q_{\delta})
& \mathcal{O}_{ld} & (\overline{l}_{\alpha}\gamma_\mu l_{\beta})(\overline{d}_{\gamma}\gamma^\mu d_{\delta})
& \mathcal{O}_{eluq}' & (\overline{e}_{\alpha}\sigma_{\mu\nu} l_{\beta}^j)\epsilon_{jk}(\overline{u}_{\gamma}\sigma^{\mu\nu} q_{\delta}^k)
\\
\midrule
 \mathcal{O}_{Ne} & (\overline{N}_{\alpha}\gamma_\mu N_{\beta})(\overline{e}_{\gamma}\gamma^\mu e_{\delta})
 & \mathcal{O}_{Nl} & (\overline{N}_{\alpha}\gamma_\mu N_{\beta})(\overline{l}_{\gamma}\gamma^\mu l_{\delta})
 & \mathcal{O}_{Nlel} & (\overline{N}_{\alpha} l_{\beta}^j)\epsilon_{jk}(\overline{e}_{\gamma} l_{\delta}^k)
\\
 \mathcal{O}_{Nu} & (\overline{N}_{\alpha}\gamma_\mu N_{\beta})(\overline{u}_{\gamma}\gamma^\mu u_{\delta})
 & \mathcal{O}_{Nq} &(\overline{N}_{\alpha}\gamma_\mu N_{\beta})(\overline{q}_{\gamma}\gamma^\mu q_{\delta})
 & \mathcal{O}_{lNqd} & (\overline{l}_{\alpha}^j N_{\beta})\epsilon_{jk}(\overline{q}_{\gamma}^k d_{\delta})
\\
\mathcal{O}_{Nd} & (\overline{N}_{\alpha}\gamma_\mu N_{\beta})(\overline{d}_{\gamma}\gamma^\mu d_{\delta})
&&& \mathcal{O}_{lNqd}' & (\overline{l}_{\alpha}^j\sigma_{\mu\nu} N_{\beta})\epsilon_{jk}(\overline{q}_{\gamma}^k\sigma^{\mu\nu} d_{\delta})
\\
\mathcal{O}_{eNud} & (\overline{e}_{\alpha}\gamma_\mu N_{\beta})(\overline{u}_{\gamma}\gamma^\mu d_{\delta})
&&& \mathcal{O}_{lNuq} & (\overline{l}_{\alpha}^j N_{\beta})(\overline{u}_{\gamma} q_{\delta}^j)
\\
\bottomrule
\tabuphantomline
\end{tabu}
\caption{Four-fermion dimension-six operators that give rise to neutrino interactions including only SM fields (upper sections) and including SM fields and sterile neutrino singlets (lower sections).}
\label{tab:4fermionSMEFT}
}
\end{table}
\begin{table}
{\centering
\begin{tabu}{X[0.5,$,c]X[4,$,c]X[$,c]X[5,$,c]X[$,c]X[6,$,c]}
\toprule
 \multicolumn2{c}{$\psi^2\varphi^3$} & \multicolumn2{c}{$\psi^2 X \varphi  $}
 & \multicolumn2{c}{$\psi^2\varphi^2$} \\
\midrule
\mathcal{O}_{e\varphi} & (\varphi^\dagger\varphi)(\overline{l}_{\alpha}e_\beta\varphi)
& \mathcal{O}_{eW} & (\overline{l}_{\alpha}\sigma^{\mu\nu}e_{\beta})\tau^I\varphi W^I_{\mu\nu}
& \mathcal{O}_{\varphi l}^{(1)} & i\left(\varphi^\dagger \dvec{D}_\mu \varphi\right)
\left(\overline l_\alpha \gamma^\mu l_\beta\right)
\\
&
& \mathcal{O}_{eB} & (\overline{l}_{\alpha}\sigma^{\mu\nu}e_{\beta})\varphi B_{\mu\nu}
& \mathcal{O}_{\varphi l}^{(3)} & i\left(\varphi^\dagger \tau^I 
\dvec{D}_\mu \varphi\right)
\left(\overline l_\alpha \tau^I \gamma^\mu l_\beta\right)
\\
\midrule
\mathcal{O}_{N\varphi} & (\varphi^\dagger\varphi)(\overline{l}_{\alpha}N_\beta\widetilde\varphi)
& \mathcal{O}_{NW} & (\overline{l}_{\alpha}\sigma^{\mu\nu}N_{\beta})\tau^I\widetilde\varphi\, W^I_{\mu\nu}
& \mathcal{O}_{\varphi N} & i\left(\varphi^\dagger \dvec{D}_\mu \varphi\right)\left(\overline N_\alpha \gamma^\mu N_\beta\right)
\\
&
& \mathcal{O}_{NB} & (\overline{l}_{\alpha}\sigma^{\mu\nu}N_{\beta})\widetilde\varphi B_{\mu\nu}
& \mathcal{O}_{\varphi Ne} & i\left(\widetilde\varphi^\dagger D_\mu \varphi\right)\left(\overline N_\alpha \gamma^\mu e_\beta\right)
\\
\bottomrule
\tabuphantomline
\end{tabu}
\caption{Fermion-boson-mixed dimension-six operators that give rise to neutrino interactions including only SM fields (upper sections) and including SM fields and sterile neutrino singlets (lower sections).}
\label{tab:bosonSMEFT}
}
\end{table}

In the cases where taking the hermitian conjugate results in only a reordering of indices, we would like to omit adding the hermitian conjugate to the given operator 
and instead impose
\[
C^{\abgd}_{X}=(C^{\beta\alpha\delta\gamma}_{X})^*
\]
for $X=ll,\, lq(1),\, lq(3),\, le,\, lu,\, ld,\, Ne,\, Nu,\, Nd,\, Nl,\, Nq$. All other four-fermion operators need explicitly added hermitian conjugates. Concerning the mixed operators, we introduced hermitian Higgs-derivative terms (following \cite{warsaw})
\[
i(\varphi^\dagger \dvec{D}_\mu \varphi) \equiv 
i\left(\varphi^\dagger D_\mu \varphi - (D_\mu\varphi)^\dagger \varphi\right),\qquad
i(\varphi^\dagger \dvec{D}^I_\mu \varphi) \equiv 
i\left(\varphi^\dagger \tau^I D_\mu \varphi - (D_\mu\varphi)^\dagger\tau^I \varphi\right),
\]
and take $\mathcal{O}_{\varphi l}^{(1)},\mathcal{O}_{\varphi l}^{(3)},\mathcal{O}_{\varphi N}$ hermitian, i.e.
\[
C_X^{\alpha\beta} = (C^{\beta\alpha}_X)^*\,,\qquad X=\varphi l(1),\,\varphi l(3),\,\varphi N.
\]

\subsection{Operator matching}
Before trying to match the GNI parameters in \eqref{eq:epsilonNC} and \eqref{eq:epsilonCC} to the Wilson coefficients of the operators of Tables \ref{tab:4fermionSMEFT} and \ref{tab:bosonSMEFT}, it is necessary to discuss the flavour basis in which the latter are considered. As discussed in \cite{Cirigliano:2009wk}, the fact that the SM gauge interactions are invariant under U(3) flavour rotations, while Yukawa matrices and Wilson coefficients of the dimension-six operators are changed, makes it possible to choose a flavour basis such that the down-type and the (SM) lepton Yukawa matrices are diagonal. In this basis, the transformation between flavour (primed) and mass (unprimed) eigenstates reads
\[
u_{L,\alpha}' = V_{\alpha\beta}^\dagger\, u_{L,\beta} \,,\quad u_{R,\alpha}' = u_{R,\alpha} \,,\quad
d_{L,\alpha}' = d_{L,\alpha} \,,\quad 
d_{R,\alpha}' = d_{R,\alpha}' \,, \quad 
e_\alpha' = e_\alpha\,,
\]
with the CKM matrix $V$ and explicit chirality indices to distinguish components of the left-handed quark doublet $q$ and right-handed singlets $u$ and $d$. This means that, expanding the operators, wherever a factor of $u_L$ appears, a factor of $V$ has to be inserted when changing to the mass basis. One needs to keep in mind that the following expressions and bounds depend on this choice of basis. Unlike in the SM, now the CKM matrix also appears in some neutral currents. For instance,
if the left-handed neutral up quark current is contracted with any matrix $C$ in flavour space, the basis change leads to
\[
C_{\alpha\beta}\,\overline{u}_{\alpha}'\gamma^\mu P_L u_\beta'
= C_{\alpha\beta}V_{\gamma\alpha} V_{\beta\delta}^\dagger\, \overline{u}_{\gamma} \gamma^\mu P_L u_\delta\,.
\]
In the SM this current is coupled to the $Z$ boson flavour-diagonally ($C_{\alpha\beta}=g_L^u\delta_{\alpha\beta}$), such that the mixing matrices cancel due their unitarity. However, for general $C$ as could be induced by SMNEFT operators they do not. 

Evaluating the four-fermion operators, we get the leptonic neutral-current contributions shown in the second column of \autoref{tab:coefficientsNC}. Note, however, that the operators $\mathcal{O}_{ll}$ and $\mathcal{O}_{le}$ appearing in the coefficients $\epsilon_{L,e}$ and $\epsilon_{R,e}$ also lead to non-standard charged-lepton interactions and are therefore subject to caution. Therefore we marked them in red and continue this notation in the following. The semileptonic neutral-current contributions are listed in the third and fourth columns, respectively.
Finally, we collect the charged-current (semileptonic) contributions in \autoref{tab:coefficientsCC}. At this stage we can already conclude that interesting interrelations of different operators to different coefficients \emph{not} involving dangerous operators are possible only in the leptonic segment
$\{\epsilon_{S/P,e},\epsilon_{T,e}\}$ and in the semileptonic segment
$
\{\epsilon_{S/P,u},\epsilon_{S/P,d},\widetilde\epsilon_{S,ud},\widetilde\epsilon_{P,ud}\}.
$
In the other cases, the correspondence of operators to coefficients is one-to-one in the basis choice of \autoref{tab:4fermionSMEFT}. However, typically a mixing between different operators can be induced when considering RG running. Therefore, the absence of an operator at a certain scale does not always imply its absence at all scales. In particular, \cite{Gonzalez-Alonso:2017iyc} find that the charged-current (pseudo)scalar and tensor GNI mix as one runs from \SI{2}{GeV} to the weak scale, while the left and right vectors do not mix. A calculation of the RG running of all the GNI and SMNEFT operators is called for, but beyond the scope of this work. Here, we will only discuss low-energy ($\lsim\SI{2}{GeV}$) bounds and note that the extension of the results to the weak scale and beyond is potentially subject to  RG running and operator mixing.

\begin{table}
{\centering
\begin{tabu}{X[$,c,.5]X[$,c]X[$,c]X[$,c]}
\toprule
  & e & u & d\\
\midrule
-\epsilon^{\abgd}_{L,f} & 
{\color{red!50!black}{C^{\abgd}_{ll}+C^{\gamma\delta\alpha\beta}_{ll}}}  
&
{\color{red!50!black}{V_{\gamma\mu} V_{\nu\delta}^\dagger\left(C^{\alpha\beta\mu\nu}_{lq(1)}+C^{\alpha\beta\mu\nu}_{lq(3)}\right)}} &
{\color{red!50!black}{C^{\abgd}_{lq(1)}-C^{\abgd}_{lq(3)}}}
\\
\addlinespace[5pt]
-\widetilde\epsilon^{\abgd}_{L,f} & 
C^{\abgd}_{Nl} &
V_{\gamma\mu} V_{\nu\delta}^\dagger C^{\alpha\beta\mu\nu}_{Nq} &
C^{\abgd}_{Nq}
\\
\addlinespace[5pt]
-\epsilon^{\abgd}_{R,f} & 
{\color{red!50!black}{C^{\abgd}_{le}}}  &
{\color{red!50!black}{C^{\abgd}_{lu}}} &
{\color{red!50!black}{C^{\abgd}_{ld}}}
\\
\addlinespace[5pt]
-\widetilde\epsilon^{\abgd}_{R,f} & 
C^{\abgd}_{Ne} &
C^{\abgd}_{Nu} &
C^{\abgd}_{Nd}
\\
\addlinespace[5pt]
-\epsilon^{\abgd}_{S,f} & \frac12C^{\abgd}_{Nlel}+\frac14C^{\gamma\beta\alpha\delta}_{Nlel} &
V_{\gamma\nu}(C^{\beta\alpha\delta\nu}_{lNuq})^* &
(C^{\beta\alpha\delta\gamma}_{lNqd})^*
\\
\addlinespace[5pt]
-\epsilon^{\abgd}_{P,f} & \frac12C^{\abgd}_{Nlel}+\frac14C^{\gamma\beta\alpha\delta}_{Nlel} &
-V_{\gamma\nu}(C^{\beta\alpha\delta\nu}_{lNuq})^* &
(C^{\beta\alpha\delta\gamma}_{lNqd})^*
\\
\addlinespace[5pt]
-\epsilon^{\abgd}_{T,f} & 
\frac18C^{\gamma\beta\alpha\delta}_{Nlel} &
0 &
(C^{'\beta\alpha\delta\gamma}_{lNqd})^*
\\
\bottomrule
\end{tabu}
\caption{Neutral-current general interaction coefficients appearing in \eqref{eq:epsilonNC} and their relation to the (dimensionless) coefficients of gauge-invariant dimension-six four-fermion operators in \autoref{tab:4fermionSMEFT}. The columns relate to $f=e,u,d$, respectively. 
The indices $\alpha, \beta = e,\mu,\tau,$ and  $\gamma, \delta, \mu, \nu = 1,2,3$ denote the generation numbers of leptons and quarks.  Interactions which also lead to new interactions among four charged fermions are printed in red. 
}
\label{tab:coefficientsNC}
}
\end{table}

\begin{table}
{\centering
\begin{tabu}{X[$,c,0.5]X[$,c]X[$,c,0.5]X[$,c]}
\toprule
-\epsilon^{\abgd}_{L,ud} & 
{\color{red!50!black}{\frac{V_{\gamma\nu}}{V_{\gamma\delta}}2C^{\alpha\beta\nu\delta}_{lq(3)}}}
&
-\widetilde\epsilon^{\abgd}_{L,ud} & 
0
\\
\addlinespace[5pt]
-\epsilon^{\abgd}_{R,ud} & 
0
  &
-\widetilde\epsilon^{\abgd}_{R,ud} & 
\frac{1}{V_{\gamma\delta}}C^{\abgd}_{eNud}
\\
\addlinespace[5pt]
-\epsilon^{\abgd}_{S,ud} & 
{\color{red!50!black}{\frac{1}{V_{\gamma\delta}}\left(V_{\gamma\nu}C^{\alpha\beta\nu\delta}_{elqd}+C^{\abgd}_{eluq}\right)}} &
-\widetilde\epsilon^{\abgd}_{S,ud} & 
\frac{1}{V_{\gamma\delta}}\left(C^{\abgd}_{lNuq}-V_{\gamma\nu} C^{\alpha\beta\nu\delta}_{lNqd}\right)
\\
\addlinespace[5pt]
-\epsilon^{\abgd}_{P,ud} & 
{\color{red!50!black}{\frac{1}{V_{\gamma\delta}}\left(-V_{\gamma\nu}C^{\alpha\beta\nu\delta}_{elqd}+C^{\abgd}_{eluq}\right)}} &
-\widetilde\epsilon^{\abgd}_{P,ud} & 
\frac{1}{V_{\gamma\delta}}\left(C^{\abgd}_{lNuq}+V_{\gamma\nu} C^{\alpha\beta\nu\delta}_{lNqd}\right)
\\
\addlinespace[5pt]
-\epsilon^{\abgd}_{T,ud} & 
{\color{red!50!black}{\frac{1}{V_{\gamma\delta}}C^{'\abgd}_{eluq}}} &
-\widetilde\epsilon^{\abgd}_{T,ud} & 
-\frac{V_{\gamma\nu}}{V_{\gamma\delta}}C^{'\alpha\beta\nu\delta}_{lNqd}
\\
\bottomrule
\end{tabu}
\caption{Charged-current general interaction coefficients appearing in \eqref{eq:epsilonCC} and their relation to the (dimensionless) coefficients of gauge-invariant dimension-six four-fermion operators in \autoref{tab:4fermionSMEFT}. 
The indices $\alpha, \beta = e,\mu,\tau,$ and  $\gamma, \delta, \mu, \nu = 1,2,3$ denote the generation numbers of leptons and quarks. 
Interactions which also lead to new interactions among four charged fermions are printed in red.}
\label{tab:coefficientsCC}
}
\end{table}

The additional charged-fermion interactions induced by the operators with coefficients marked in red above can be summarised as
\[
\label{eq:charged-fermion-ints}
\begin{split}
\lagr_{} &= \sqrt{8}G_F\left\{
(C_{ll}^{\abgd}+C_{ll}^{\gamma\delta\alpha\beta})(\overline{e}_\alpha\gamma_\mu P_Le_\beta)(\overline{e}_{\gamma}\gamma^\mu P_Le_{\delta})
+C_{le}^{\abgd}(\overline{e}_\alpha\gamma_\mu P_Le_\beta)(\overline{e}_{\gamma}\gamma^\mu P_Re_{\delta})\right.
\\
&
\quad +V_{\gamma\mu} V_{\nu\delta}^\dagger (C^{\alpha\beta\mu\nu}_{lq(1)}-C^{\alpha\beta\mu\nu}_{lq(3)})(\overline{e}_\alpha\gamma_\mu P_Le_\beta)(\overline{u}_{\gamma}\gamma^\mu P_L u_{\delta})+C_{lu}^{\abgd}(\overline{e}_\alpha\gamma_\mu P_Le_\beta)(\overline{u}_{\gamma}\gamma^\mu P_Ru_{\delta})\\
&\quad+(C_{lq(1)}^{\abgd}+C_{lq(3)}^{\abgd})(\overline{e}_\alpha\gamma_\mu P_Le_\beta)(\overline{d}_{\gamma}\gamma^\mu P_Ld_{\delta})
+C_{ld}^{\abgd}(\overline{e}_\alpha\gamma_\mu P_Le_\beta)(\overline{d}_{\gamma}\gamma^\mu P_Rd_{\delta})\\
&\quad-V^\dagger_{\nu\delta}C_{eluq}^{\alpha\beta\gamma\nu}(\overline{e}_\alpha P_Le_\beta)(\overline{u}_{\gamma} P_Lu_{\delta})
-V^\dagger_{\nu\delta}C_{eluq}^{'\alpha\beta\gamma\nu}(\overline{e}_\alpha \sigma_{\mu\nu} P_Le_\beta)(\overline{u}_{\gamma} \sigma^{\mu\nu} P_Lu_{\delta})\\
&\quad
\left.+\, C_{elqd}^{\abgd}(\overline{e}_\alpha P_Le_\beta)(\overline{d}_{\gamma} P_Rd_{\delta})\right\}.
\end{split}
\]
We will discuss, for some examples, how the constraints on these charged-fermion interactions can be used to place strong indirect bounds on neutrino interactions from gauge-invariant operators in the next section.

We turn to the mixed operators in \autoref{tab:bosonSMEFT}. Below the electroweak scale, the operators of the form $\psi^2\varphi^2$ reduce to corrections to Dirac mass terms of charged leptons and neutrinos and are therefore uninteresting for our purposes. Up next, the $\psi^2 X\varphi$-type operators modify the couplings of neutrinos and charged leptons to the $W$ and $Z$ bosons, as well as the photon. Unless some very precise cancellations appear, between e.g.\  $C_{eW}$ and $C_{eB}$, this leads, for instance, to deviations in QED, which are extremely constrained. 
Therefore, from the viewpoint of sizable neutrino interactions, it is reasonable to focus on the operators of type $\psi^2\varphi^2$ which below the weak scale lead to the modification of the SM couplings to the weak gauge bosons. 
We skip the full description and refer to \autoref{sec:mixedoperators} for details, since in this study we focus on the four-fermion operators. Let us, however, summarise the key points. Operators $\mathcal{O}_{\varphi N}$ and $\mathcal{O}_{\varphi Ne}$ lead to interactions which are absent in the SM, namely a $Z$ coupling to right-handed neutrinos, and a right-handed leptonic coupling to $W$. On the other hand, $\mathcal{O}_{\varphi l(1)}$ modifies the left-handed $Z$ coupling of neutrinos and charged leptons, while $\mathcal{O}_{\varphi l(3)}$ additionally modifies the leptonic $W$ coupling, both of which are related to the experimental determination of fundamental SM parameters. Therefore, a detailed discussion would need to account for the determination of SM parameters in the presence of new physics.

Let us conclude this section by making a few comments about the implications of our results. First of all, we note that some GNI are impossible to embed in SMNEFT with four-fermion operators
up to dimension six,
 namely neutral-current tensor interactions with up-type quarks, and the charged-current couplings $\epsilon_{R,ud}$ and $\widetilde\epsilon_{L,ud}$. Secondly, and contrary to frequently made assumptions, in general we should expect that the neutral-current interactions of neutrinos with up-type quarks are different from those with down-type quarks as becomes clear from \autoref{tab:coefficientsNC}. Finally, since the NSI coefficients $\epsilon_L$ and $\epsilon_R$ are either zero or accompanied by charged-fermion interactions of the same strength, they cannot be large (close to the weak scale) from a high-energy EFT perspective. In this respect, those GNI involving right-handed neutrinos are favoured, however requiring to go beyond SMEFT by introducing at least either a right-handed Dirac partner for a SM neutrino or a light sterile Majorana neutrino. The only ways to circumvent these conclusions are to either go to higher order in the EFT expansion, or abandon the premiss of the high-energy origin of new physics and introduce new low-energy physics that non-linearly realises the SM symmetries.

\section{Phenomenology}
\label{sec:pheno}
In this section, we investigate some phenomenological implications of the assumption that new neutrino interactions are induced by the gauge-invariant dimension-six operators introduced in the previous section. Considering examples, we demonstrate how the new interactions induced by the ``dangerous'' operators involving also new charged-fermion interactions generally have to be very small (much below near-future sensitivities) to satisfy current bounds. However, the operators involving light right-handed neutrinos can lead to detectable effects.  Note that here we do not consider renormalisation group (RG) running of Wilson coefficients as would be necessary when considering observables at different energy scales. Instead, we restrict ourselves to compare low-energy ($\lsim\SI{2}{GeV}$) measurements, for instance, beta decays and coherent elastic neutrino-nucleus scattering (CE$\nu$NS). This is well-motivated, since decay and low-energy experiments in general are more sensitive to light (keV and below) sterile neutrinos, whereas heavier (GeV--\SI{500}{GeV}) sterile neutrinos are being searched for at the LHC \cite{Aad:2019kiz,Sirunyan:2018mtv,Aad:2015xaa,Chatrchyan:2012fla}. 
There do exist, however, collider bounds on contact interactions, some of which we address for comparison.
Here, we will not consider reactor neutrino oscillation experiments as another source of low-energy data, but refer to \cite{Falkowski:2019xoe} for a recent analysis in the context of SMEFT.

\subsection{Implications of charged lepton flavour violation searches}
\label{sec:pheno-clfv}
Several experiments are dedicated to the search for the transition from muon to electron flavour in processes like $\mu\rightarrow eee$, $\mu\rightarrow e\gamma$, or $\mu \rightarrow e$ conversion in nuclei \cite{DEGOUVEA201375,Calibbi:2017uvl}. The maximal size of corresponding neutrino interactions changing from muon to electron flavour is given by the experimental bounds on charged lepton flavour conversion under the assumption that both are induced by the dimension-six operators of \autoref{tab:4fermionSMEFT}. Thereby we assume contributions from operators of dimension $d>6$ to be subdominant. The current single-parameter bounds (90\% CL) on the relevant subset of the dimension-six operators are discussed in the following. We summarise them in \autoref{tab:clfvbounds} with respect to GNI coefficients, and in \autoref{tab:eftbounds} with respect to SMNEFT coefficients.

\begin{itemize}

\item $\mu\rightarrow eee$: The relevant coefficients are ${C}_{ll}^{\mu eee}$, $C_{le}^{\mu eee}$ and ${C}_{le}^{ee\mu e}$ and constrained to be at most of order $10^{-6}$ \cite{Calibbi:2017uvl}\footnote{Values given therein for $C_a$ at $\Lambda=\SI{1}{\TeV}$ translate to ours by a factor of $(\sqrt{8}G_F\si{TeV^2})^{-1}\approx 1/33$, since we normalised our coefficients with respect to the weak scale.
} by the results of SINDRUM \cite{sindrum}. Further improvements by up to two orders of magnitude are expected with the upcoming Mu3e experiment\cite{mu3e,mu3e2}, noting that the branching ratio is proportional to the operator coefficients squared.
This compares to the current direct bounds on NSI of the order $10^{-1}$ on the related coefficients $\epsilon_{L,e}^{\mu eee}$ and $\epsilon_{R,e}^{\mu eee}$ from $\nu_\mu e$ scattering in CHARM-II \cite{Barranco:2007ej,charmii,charmiib}. Improvements are expected by about an order of magnitude from $\nu_\mu e$ scattering data at the DUNE near detector \cite{Bischer:2018zcz}. There are no bounds implied by $\mu\rightarrow eee$ on the scalar and tensor interactions $\epsilon_{S/P/T,e}^{e\mu ee}$ (cf.\ \autoref{tab:coefficientsNC}), but the direct bounds
from CHARM-II calculated in \cite{Rodejohann:2017vup} can be applied, namely
$|\epsilon_{T,e}^{\alpha\mu ee}|\leq 0.04$, $|\epsilon_{S/P,e}^{\alpha\mu ee}|\leq0.4$ for any single $\alpha$, which is also expected to improve with DUNE near detector data by a factor of 2-4 \cite{Bischer:2018zcz}.

\item $\mu\rightarrow e$ in nuclei: The relevant vector-like operators $C^{e\mu11}_{ld,lu}$ have been constrained in \cite{Feruglio:2015gka} to  the order $10^{-8}$ employing results of SINDRUM-II \cite{sindrum2} on the rate of $\mu^-\mathrm{Au}\rightarrow e^-\mathrm{Au}$. 
Extending the results of \cite{Feruglio:2015gka}, we apply the formulae of \cite{Kitano:2002mt} (Method 1) to constrain also $C_{lq(1)}$, $C_{lq(3)}$, and the coefficients of scalar operators $C_{elqd}$ and $C_{eluq}$ with index structures $e\mu 11$ to all be of order $10^{-7}$. Even stronger bounds are expected from the Mu2e experiment \cite{mu2e}. This compares to direct bounds on $\epsilon_{L,q}^{e\mu11}$ and $\epsilon_{R,q}^{e\mu11}$ at the order of $10^{-2}$ from evaluating neutrino-nucleon scattering data of CHARM and CDHS \cite{charm,Blondel:1989ev,Farzan:2017xzy,cirigliano2}. The scalar charged-current interactions can be most stringently constrained from pion decay, $|\mathrm{Re}(\epsilon_{S,ud}^{e\mu11})|\lsim10^{-3}$, $|\mathrm{Re}(\epsilon_{P,ud}^{e\mu11})|\lsim10^{-4}$, where the difference is due to the ratio of $\pi\rightarrow e\nu (\gamma)$ to $\pi\rightarrow \mu\nu (\gamma)$ being very sensitive to pseudoscalar and axial couplings  \cite{cirigliano2}.

\item $\mu\rightarrow e\gamma$: Through a one-loop diagram, the coefficients $C_{le}^{\mu\mu\mu e,e\mu\mu\mu}$ and $C_{le}^{\mu\tau\tau e,e\tau\tau\mu}$ contribute to this decay channel. The bounds on these coefficients from MEG \cite{TheMEG:2016wtm} are of the order $10^{-6}$-$10^{-7}$ \cite{Calibbi:2017uvl}. The related $\epsilon$-coefficients involve NSI with muons and taus instead of electrons. We are not aware of a study of quantitative bounds on this type of interaction.

\end{itemize}
If we consider operators with tau flavour indices, however, for instance tau decays like $\tau\rightarrow\mu\mu\mu$ and $\tau\rightarrow eee$, constraints are somewhat weaker, but still beyond the typical sensitivity of neutrino experiments:
\begin{itemize}

\item $\tau\rightarrow\mu\mu\mu$: The relevant coefficients are ${C}_{ll}^{\tau\mu\mu\mu}$, $C_{le}^{\mu\tau\mu\mu}$, $C_{le}^{\mu\mu\mu\tau}$ and constrained to be at most of order $10^{-4}$ by the results of Belle \cite{Calibbi:2017uvl,Hayasaka:2010np}. These operators would give rise to NSI of neutrinos with muons and taus.

\item $\tau\rightarrow eee$: The relevant coefficients are ${C}_{ll}^{\tau eee}$, $C_{le}^{\tau eee}$, $C_{le}^{ee\tau e}$ and constrained to be at most of order $10^{-4}$ also by the results of Belle \cite{Calibbi:2017uvl,Hayasaka:2010np}. These give rise, in particular, to tau-electron flavour-changing NSI with electrons, $\epsilon_{L,e}^{\tau eee}$, $\epsilon_{R,e}^{\tau eee}$. Direct bounds on such interactions are at the order of $10^{-1}$ coming from $e^+e^-\rightarrow\overline\nu\nu\gamma$ data of LEP \cite{Barranco:2007ej}.

\end{itemize}

In summary, dimension-six operators that can lead to lepton flavour changing NSI of the type $\epsilon_{L/R}$ and exotic charged-current interactions of the type $\epsilon_{S/P/T,ud}$ of left-handed neutrinos with the first generations of charged fermions are highly constrained.
We conclude that in neutrino experiments the prospects of detecting exotic (pseudo)scalar or tensor interactions $\epsilon_{S/P/T}$ are better in the sense that less of the parameter space is already ruled out by neutrino-independent measurements. 
This conclusion relies on the invoked high-energy origin, and, conversely, the detection of such interactions would strongly hint towards more exotic new physics scenarios. 
Let us stress that in the derivation of these bounds, we did not require any particular flavour structure, such that a given bound only holds for the specified flavour indices.
Assuming that there is no flavour structure or hierarchy would imply that operators involving $\tau$ leptons are impossible to observe, as in this case the strong limits from $\mu$-$e$ transitions would apply for those as well. A flavour symmetry would be a candidate to generate a hierarchy among the different Wilson coefficients.

\begin{table}
{\centering
\begin{tabu}{X[$,c]X[$,c]X[$,c]X[$,c]}
\toprule
|\epsilon^{e\mu ee}|\text{ or }|\epsilon^{e\mu 11}| &\mathrm{Direct} & \mathrm{CLFV} &\text{Operators}\\
\midrule
\epsilon_{L,e} & \num{1.3e-1}\text{\cite{Barranco:2007ej}}& \num{1.4e-6} & \mathcal{O}_{ll}
\\
\epsilon_{R,e} & \num{1.3e-1}\text{\cite{Barranco:2007ej}}& \num{1.0e-6} & \mathcal{O}_{le}
\\
\epsilon_{L,u} &  \num{2.3e-2}\text{\cite{Escrihuela:2011cf}}& \num{3.3e-7} & \mathcal{O}_{lq(1)},\mathcal{O}_{lq(3)}
\\
\epsilon_{L,d} & \num{2.3e-2}\text{\cite{Escrihuela:2011cf}}& \num{3.3e-7} & \mathcal{O}_{lq(1)},\mathcal{O}_{lq(3)}
\\
\epsilon_{R,u} & \num{3.6e-2}\text{\cite{Escrihuela:2011cf}} & \num{6.0e-8} & \mathcal{O}_{lu}
\\
\epsilon_{R,d} & \num{3.6e-2}\text{\cite{Escrihuela:2011cf}} & \num{5.3e-8} & \mathcal{O}_{ld}
\\
\epsilon_{L,ud} & \num{2.6e-2}\text{\cite{Biggio:2009nt}}  & \num{6.6e-7} & \mathcal{O}_{lq(3)}
\\
\mathrm{Re}(\epsilon_{S,ud}) & \num{8e-3}\text{\cite{cirigliano2}} & \num{3.0e-8} & \mathcal{O}_{elqd},\mathcal{O}_{eluq}
\\
\mathrm{Re}(\epsilon_{P,ud}) & \num{4e-4}\text{\cite{cirigliano2}} &\num{3.0e-8} & \mathcal{O}_{elqd},\mathcal{O}_{eluq}
\\
\bottomrule
\end{tabu}
\caption{Single-parameter bounds (90\% CL) on general interaction parameters that are related to CLFV from muon to electron flavour. The second column lists direct bounds from neutrino experiments, while the third column features bounds from CLFV searches under the assumption that the interactions are induced by one of the operators in the fourth column.
}
\label{tab:clfvbounds}
}
\end{table}

For completeness, let us discuss results on flavour-\emph{conserving} coefficients in SMEFT.
Note that when interpreting the SM as an EFT, it is in general not sufficient to fix SM parameters from measurements and subsequently study the effect of higher-order operators. Instead, if a SM parameter is determined from a certain physical process that also has a contribution from new operators, it needs to be rescaled to include the corresponding Wilson coefficient. As a famous example, the Fermi constant $G_F$ is usually determined by muon decay and as such would include a contribution from $\mathcal{O}_{ll}^{e\mu\mu e}$. 
Since the weak interactions respect lepton flavour, flavour-diagonal operators are especially prone to being absorbed into definitions of SM parameters, requiring a global analysis. Such an analysis of all flavour-diagonal SMEFT four-fermion operators was carried out in \cite{Falkowski:2017pss}, to which we refer for details.\footnote{This study does not yet include the CE$\nu$NS results of the COHERENT experiment \cite{coherent}. } 
Concerning neutrino interactions with charged leptons, their results imply typically order $10^{-2}$ bounds, while the strength of neutrino-quark interactions ranges between $10^{-4}$ and $10^{-2}$.
The flavour-diagonal interactions can also be probed with $pp\rightarrow ll$ processes at the LHC, albeit at much higher energies \cite{Aaboud:2017buh,Sirunyan:2018ipj}. Both ATLAS and CMS present strong limits on vector-like contact interactions of the form of lines 2 and 3 of \eqref{eq:charged-fermion-ints}. Hence they constrain $C_{lq(1)},C_{lq(3)},C_{lu},C_{ld}$. For these particular operators, the scale of new interactions in the dielectron and dimuon channels is constrained to be larger than 20--\SI{26}{TeV} corresponding to Wilson coefficients of order $10^{-4}$. Those sensitivities generally surpass the low-energy bounds derived in \cite{Falkowski:2017pss}. However, the comparison requires that the EFT description be still valid at energy scales of \SI{13}{TeV}.

\subsection{Exotic interactions in coherent scattering and beta decay}
\label{sec:pheno-exotic}
Based on the observation that interesting correlations between the sets of coefficients $
\{\epsilon_{S/P,u},\epsilon_{S/P,d},\widetilde\epsilon_{S,ud},\widetilde\epsilon_{P,ud}\}
$ and 
$
\{\epsilon_{T,d},\widetilde\epsilon_{T,ud}
\}
$
exist, involving the operators $\mathcal{O}_{lNuq}$ and $\mathcal{O}_{lNqd}$, and $\mathcal{O}_{lNqd}'$ respectively, we investigate the prospects of observing such correlations between different experiments. Note that by analogy to the operator structure $\epsilon_{S,ud}$, $\epsilon_{P,ud}$, $\epsilon_{T,ud}$ considered in \cite{Gonzalez-Alonso:2018omy}, we would expect a mixing of the GNI parameters when considering RG running although to our knowledge no calculation of those effects has been performed.\footnote{
Interestingly, above the weak scale $\mathcal{O}_{elqd}$ does not mix with $\mathcal{O}_{eluq}$ and $\mathcal{O}_{eluq}'$, which mix among each other. We expect a similar situation for $\mathcal{O}_{lNuq}$, $\mathcal{O}_{lNqd}$, and $\mathcal{O}_{lNqd}'$, where the latter two are connected by the fact that they would mix in a Fierz-transformed operator basis.
}
Clearly, the charged-current operators in \eqref{eq:epsilonCC} would have an effect on beta decay \cite{Cirigliano:2012ab,Ludl:2016ane,Gonzalez-Alonso:2018omy}. Taking first-generation indices on the quark side, electron index for the charged lepton, and only the scalar coefficients, one has
\[\label{eq:ccpheno}
\Delta\lagr^\mathrm{CC}=-\frac{G_FV_{ud}}{\sqrt{2}}\sum_{\beta}\left(\widetilde\epsilon_{S,ud}^{e\beta11}  (\overline e \projr N_\beta)(\overline u d)
-\widetilde\epsilon_{P,ud}^{e\beta11} (\overline e \projr N_\beta)(\overline u\gamma^5d)
\right)\plushc
\]
This expression would be relevant not only for nuclear beta decays but also for charged pion decay which indeed yields the strongest direct bound on $\widetilde\epsilon_{P,ud}$, as expanded below.  

Excitingly, a new way to probe the neutral-current operators, has recently become feasible, namely coherent elastic neutrino-nucleus scattering (CE$\nu$NS) \cite{Freedman:1973yd} which was successfully observed by COHERENT \cite{coherent} in the partially coherent regime. 
Reactor neutrinos are a very interesting source, since they allow to measure this process in the fully coherent regime, as currently pursued for instance by the CONUS collaboration \cite{maneschg_werner_2018_1286927}.
In the context of reactor neutrinos (electron antineutrinos) the sensitivity is also towards first generations,
\[\label{eq:ncpheno}
\Delta\lagr^\mathrm{NC}=-\frac{G_F}{\sqrt{2}}\sum_{\substack{\beta\\ \psi=u,d}}\left(\epsilon_{S,\psi}^{\beta e11}  (\overline N_\beta \projl \nu_e)(\overline \psi \psi)
-\epsilon_{P,\psi}^{\beta e11} (\overline N_\beta \projl \nu_e)(\overline \psi\gamma^5\psi)
\right)\plushc
\]
The impact of general interactions on the CE$\nu$NS cross-sections has been calculated in \cite{Lindner:2016wff}. The result of only the (pseudo)scalar and tensor sectors reads
\[
\frac{\der \sigma}{\der T} = \frac{G_F^2M}{4\pi}\left(\xi_S^2\frac{MT}{2E_\nu^2} + \xi^2_T \left(1 - \frac{T}{T_\mathrm{max}} + \frac{MT}{4E_\nu^2}\right)\right),
\] 
where $M$ and $T$ denote the mass and recoil energy of the nucleus, $E_\nu$ the neutrino energy, and we neglected terms of higher order in $T/E_\nu$.
For a nucleus of atomic mass $A$ and atomic number $Z$, the parameter $\xi_S$ is given by
\[
\xi^2_S = (C_S^2+D_P^2)\,,
\]
with \cite{AristizabalSierra:2018eqm}
\[\label{eq:csdef}
\begin{split}
C_S&\equiv\sum_{q=u,d}C_S^{(q)}\left[
N\frac{m_n}{m_q}f^n_{Tq}F_n(Q^2)+Z\frac{m_p}{m_q}f^p_{Tq}F_p(Q^2)
\right],
\end{split}
\]
where $N=A-Z$ is the neutron number, $F_n$ and $F_p$ are the appropriate form factors of neutron and proton (typically assumed to be equal) as functions of energy transfer $Q$, and $f^n_{Tq}$ and $f^p_{Tq}$ are related to the fraction of nucleon mass the given quark type contributes and calculated in chiral perturbation theory \cite{HAIYANGCHENG1989347}. 
The interaction parameters $C_j^{(q)}$, $D_j^{(q)}$ are a simple reparametrisation of the $\epsilon_{j,q}$, $\widetilde\epsilon_{j,q}$, given by \hyperref[eq:epsilontocd]{Equation}~\eqref{eq:epsilontocd} in \autoref{sec:details}. In particular,
\[\label{eq:epstocdshort}
C^{(q)}_S=\epsilon_{S,q}+\widetilde\epsilon_{S,q}\,,\qquad
D^{(q)}_P=i(\widetilde\epsilon_{S,q}-\epsilon_{S,q})\,.
\]
The definition of $D_P$ is the same as \eqref{eq:csdef}, with $C_S^{(q)}$ replaced by $D_P^{(q)}$. Since we would like to consider the results of \cite{AristizabalSierra:2018eqm}, we use the same values for the parameters 
\[
\begin{split}
f^p_{Tu}&=0.019\,,\\
f^n_{Tu}&=0.023\,,
\end{split}
\qquad
\begin{split}
f^p_{Td}&=0.041\,,\\
f^n_{Td}&=0.034\,,
\end{split}
\]
taken from \cite{Jungman:1995df}. 

The tensor parameter $\xi_T$ is given by
\[\label{eq:xitensor}
\xi^2_T = 8(C_T^2+D_T^2)\,,
\]
where 
\[
\label{eq:ctensor}
C_T = N(\delta^n_u C^u_T + \delta^n_d C^d_T)F_n(Q^2)
  + Z(\delta^p_u C^u_T + \delta^p_d C^d_T ) F_p(Q^2)\,,
\]
with, again, fundamental interaction parameters $C^q_T$, $D^q_T$ related to $\epsilon_T$, $\widetilde\epsilon_T$ in \eqref{eq:epsilontocd}. 
We neglect the effect of $D_T$ since it is related to the spin-dependent part of the cross-section related to $\epsilon_T - \widetilde\epsilon_T$, and use for the tensor charges $\delta^n_q$, $\delta^p_q$ the same values as in \cite{AristizabalSierra:2018eqm}, namely
\[
\begin{split}
\delta^p_{u}&=0.54 
\,,\\
\delta^n_{u}&=-0.23 
\,,
\end{split}
\qquad
\begin{split}
\delta^p_{d}&=-0.23\,,\\
\delta^n_{d}&=0.54\,,
\end{split}
\]
taken from \cite{Anselmino:2008jk}.
We remark that there are relatively large uncertainties on these parameters, such that the values we extract below should be interpreted as estimates.

In the following, we compare the sensitivity of beta decay experiments on the interactions \eqref{eq:ccpheno} to the sensitivity of CE$\nu$NS experiments on the interactions in \eqref{eq:ncpheno}, assuming they are induced by the gauge invariant operators $\mathcal{O}_{lNuq}$, $\mathcal{O}_{lNqd}$, and $\mathcal{O}_{lNqd}'$. Since the flavour of the outgoing neutrino is not measured in both cases, one may as well take $\beta=e$ here, assuming Dirac neutrinos.\footnote{More generally, $\beta$ could take any number of indices of light right-handed neutrinos beyond the SM and we would be sensitive only to the sum.} 
We consider three scenarios, first the minimal scenario of one non-vanishing scalar operator, second the scenario of both scalar operators non-vanishing and correlated, and third the scenario of only the tensor operator $\mathcal{O}_{lNqd}'$.

\subsubsection{Single operator $lNuq$}
\label{sec:pheno-singlescalar}
If we consider only $C_{lNuq}^{ee11}$ non-vanishing, the following simplified relations are derived from \hyperref[tab:coefficientsNC]{Tables}~\ref{tab:coefficientsNC} and \ref{tab:coefficientsCC}:
\[\label{eq:one-operator}
\widetilde\epsilon_{S,ud}^{ee11}=\widetilde\epsilon_{P,ud}^{ee11}
= -\frac{1}{V_{ud}}C^{ee11}_{lNuq} = \frac{1}{|V_{ud}|^2}(\epsilon_{S,u}^{ee11})^*=-\frac{1}{|V_{ud}|^2}(\epsilon_{P,u}^{ee11})^*\,.
\]
To simplify notation, in the following we will drop the $ee11$-superscript.
Using \eqref{eq:one-operator} and assuming $F_n=F_p=F$, \eqref{eq:csdef} simplifies to
\[
\begin{split}
C_S&=C_S^{(u)}F(Q^2)\left[
N\frac{m_n}{m_u}f^n_{Tu}+Z\frac{m_p}{m_u}f^p_{Tu}
\right],
\end{split}
\]
and likewise for $D_P^{(u)}$.
Since generally $\epsilon_{S,u}^{ee11}=(\widetilde{\epsilon}_{S,u}^{ee11})^*$, we can use \eqref{eq:epstocdshort} to find
\[
C_S^{(u)} = 2\,\mathrm{Re}(\epsilon_{S,u}^{ee11})\,,\qquad
D_P^{(u)} = 2\,\mathrm{Im}(\epsilon_{S,u}^{ee11})\,.
\]
Bounds from the COHERENT experiment have been derived in \cite{AristizabalSierra:2018eqm} assuming equal Helm form factors for proton and neutron. It has been pointed out that this assumption introduces rather large uncertainties for momentum transfers $Q\geq\SI{20}{MeV}$ as is the case of COHERENT, while reactor neutrino experiments such as CONUS are safe from such large uncertainties \cite{AristizabalSierra:2019zmy}.
Neglecting those for now, we can estimate the current bound on $\epsilon_{S,u}^{ee11}$ from the bound on $\xi_S$. From \cite{AristizabalSierra:2018eqm} we quote 
\[\label{eq:xisquarebound}
\begin{split}
|\xi_S|/NF(Q^2)&\leq 0.62 \quad\text{at 90\% CL,}\,\\
|\xi_S|/NF(Q^2)&\leq 1.065\quad\text{at 99\% CL,}\,
\end{split}
\]
employing CsI as target material and assuming no lepton flavour dependence, i.e.\ $\epsilon^{ee11}=\epsilon^{\mu\mu11}$. Taking $N=77.9$, $Z=55$ for Caesium nuclei, and $N=73.9$, $Z=53$ for Iodine nuclei, we obtain
\[
\begin{split}
\frac{\xi_S^2}{N^2F^2(Q^2)}
&=\sum_{i=\mathrm{Cs},\mathrm{I}}\frac{1}{N_i^2}\left((C_S^{(u)})^2+(D_P^{(u)})^2\right)\left[
N_i\frac{m_n}{m_u}f^n_{Tu}+Z_i\frac{m_p}{m_u}f^p_{Tu}
\right]^2\\
&=4|\epsilon_{S,u}^{ee11}|^2\,\sum_{i=\mathrm{Cs},\mathrm{I}}\frac{1}{N_i^2}\left[
N_i\frac{m_n}{m_u}f^n_{Tu}+Z_i\frac{m_p}{m_u}f^p_{Tu}
\right]^2,
\end{split}
\]
which amounts to
\[
\begin{split}
|\epsilon_{S,u}^{ee11}|&\leq 0.015\quad\text{at 90\% CL,}\,\\
|\epsilon_{S,u}^{ee11}|&\leq 0.026\quad\text{at 99\% CL.}\,
\end{split}
\]
Note that relaxing the assumption of no lepton flavour dependence would weaken this bound since the majority of neutrinos in the COHERENT beam are of muon flavour \cite{AristizabalSierra:2018eqm}. This would not be the case for the electron-flavoured reactor neutrinos. Indeed, as an outlook to the near future, we consider that in \cite{Lindner:2016wff}, future bounds are projected to be $\xi_S/NF\leq 0.21$ at $3\sigma$ for a Germanium target and reactor neutrinos. This translates into
\[
\begin{split}
(0.21)^2\geq \xi_S^2
&=4|\epsilon_{S,u}^{ee11}|^2F^2(Q^2)\left[
N\frac{m_n}{m_u}f^n_{Tu}+Z\frac{m_p}{m_u}f^p_{Tu}
\right]^2,
\end{split}
\]
and we take also $N=40.6$, $Z=32$ for Germanium, as well as $F^2(Q)=1$, which is sufficiently accurate for reactor neutrino energies and \si{keV}-scale energy transfers. We find
\[
|\epsilon_{S,u}^{ee11}|\leq \num{2e-4}\,.
\]
Next we would like to compare this to the bounds on charged-current interactions. 
We quote the low-energy bounds (\SI{2}{GeV}) of
\[\label{eq:tildebounds}
\begin{split}
|\mathrm{Re}(\widetilde\epsilon_{P,ud})|,|\mathrm{Im}(\widetilde\epsilon_{P,ud})|&\leq \num{2.8e-4}\,\\
|\widetilde\epsilon_{S,ud}|&\leq \num{6.3e-2}\,\\
\end{split}
\quad
\begin{split}
\text{\cite{cirigliano2}},&\\
\text{\cite{Gonzalez-Alonso:2018omy}},&
\end{split}
\]
at 90\% CL, 
where the first bound is obtained from the aforementioned ratio between the rates of pion decay to electron and to muon \cite{cirigliano2}, and the second bound is taken from a recent global fit of nuclear and neutron beta decay data \cite{Gonzalez-Alonso:2018omy}. 

The bounds implied for the Wilson coefficient $C_{lNuq}$ only differ by the mixing matrix correction $|V_{ud}|=0.97420$ \cite{Tanabashi:2018oca}, namely
\[
\begin{split}\label{eq:pheno-single-bound}
|C_{lNuq}| \leq \num{1.5e-2} \qquad &\text{from CE$\nu$NS,} \\
|C_{lNuq}| \leq \num{3.9e-4} \qquad &\text{from beta decay},
\end{split}
\]
at 90\% CL, where we quadratically added the bounds on real and imaginary parts \eqref{eq:tildebounds} in the second line.
So unfortunately with only $\mathcal{O}_{lNuq}$, the Wilson coefficient responsible for $\epsilon_{S,u}$, $\epsilon_{P,u}$ is constrained to be too small as to be probed in the near future in CE$\nu$NS. According to \autoref{tab:coefficientsCC}, this could be alleviated in the case that also $\mathcal{O}_{lNqd}$ is present with a Wilson coefficient of opposite sign, but similar magnitude.  Indeed, we would conclude
\[\label{eq:two-op-correlation}
C_{lNuq}\approx -V_{ud} C_{lNqd}\quad\Rightarrow\quad
\widetilde\epsilon_{P,ud}\approx 0\,,\qquad
\widetilde\epsilon_{S,ud}\approx -\frac{2}{V_{ud}}C_{lNuq}\,.
\]
One should keep in mind that such a relation, if true at one scale, is expected to be violated at other scales when considering RG running.
\hyperref[eq:two-op-correlation]{Equation}~\eqref{eq:two-op-correlation} would imply $\epsilon_{S,d}=\epsilon_{P,d}$ to be non-vanishing and of the same magnitude, a case which we cover in the next subsection.

\subsubsection{Correlated operators $lNuq$ and $lNqd$}
\label{sec:pheno-twoscalars}
As discussed in the previous section, a large ($\gsim10^{-4}$) magnitude of $\widetilde\epsilon_{P,ud}^{ee11}$ is excluded. Hence $C_{lNuq}^{ee11}$ must be small, unless the relation \eqref{eq:two-op-correlation} holds. In this case, we would conclude that
\[\label{eq:correlOperators}
(\widetilde\epsilon^{ee11}_{S,ud})^*\approx -\frac{2}{V_{ud}^*}(C^{ee11}_{lNuq})^*=\frac{2}{|V_{ud}|^2} \epsilon^{ee11}_{S,u}=-\frac{2}{|V_{ud}|^2}\epsilon^{ee11}_{P,u}
=-2\epsilon^{ee11}_{S,d} = -2\epsilon^{ee11}_{P,d}\,.
\]
We will again drop the $ee11$ superscript. 
This situation should be compared to the new composition of $\xi_S$ in coherent scattering. In an analogous calculation to the previous subsection, we use
\[
\begin{split}
C_S^{(u)}&=2\,\mathrm{Re}(\epsilon_{S,u})=-2\,\mathrm{Re}(V_{ud}C_{lNuq})\,,\\
C_S^{(d)}&=2\,\mathrm{Re}(\epsilon_{S,d})=2\,\mathrm{Re}(V_{ud}^{-1}C_{lNuq})\,,
\end{split}
\quad
\begin{split}
D_P^{(u)}&=2\,\mathrm{Im}(\epsilon_{S,u})=2\,\mathrm{Im}(V_{ud}C_{lNuq})\,,\\
D_P^{(d)}&=2\,\mathrm{Im}(\epsilon_{S,d})=-2\,\mathrm{Im}(V_{ud}^{-1}C_{lNuq})\,,
\end{split}
\]
and \eqref{eq:xisquarebound} to obtain
\[
\begin{split}
|C_{lNuq}| \leq \num{1.8e-2} \qquad &\text{from CE$\nu$NS,}\\
|C_{lNuq}| \leq \num{3.7e-2} \qquad &\text{from beta decay,}
\end{split}
\]
at 90\% CL, where in the second line we also translated the relevant beta decay bounds from \cite{Gonzalez-Alonso:2018omy} as quoted in \eqref{eq:tildebounds}.
This shows that bounds from beta decay and coherent scattering are currently at the same order of magnitude, assuming the scalar-only charged-current scenario. We estimate the discovery potential at a near-future Germanium-based reactor experiment as in the previous section by applying the $3\sigma$-bound of $\xi_S\leq 0.21$ from \cite{Lindner:2016wff} to the case at hand, \eqref{eq:correlOperators}, and find
\[
|C_{lNuq}|=|V_{ud}^{-1}\epsilon_{S,u}|=|V_{ud}\,\epsilon_{S,d}|\leq\num{2.0e-4}\,.
\]
We conclude that there is potential to discover those interactions in future beta decay surveys and coherent scattering experiments, and that if both are found, that would constitute a clear hint on the operator structure composed of both $\mathcal{O}_{lNuq}$ and $\mathcal{O}_{lNqd}$.

\subsubsection{Single operator $lNqd'$}
\label{sec:pheno-tensor}
Analogously to \autoref{sec:pheno-singlescalar}, we assume only $C_{lNqd}^{'ee11}$ to be non-vanishing, which leads to 
\[
\epsilon_{T,d}^{ee11} = - \widetilde\epsilon_{T,ud}^{ee11} = 
  -\frac{1}{2} C_{T,ud}^{ee11} = -(C_{lNqd}^{'ee11})^*.
\]
Since there are now only neutral-current interactions with down quarks, \eqref{eq:xitensor} reduces to
\[
\xi_T^2 = 8C_T^2 = 
8 F^2(Q^2) (C_T^d)^2 (N\delta^n_d + Z \delta^p_d)^2.
\]
Using \eqref{eq:epsilontocd} together with \eqref{eq:epsilonconstraints}, we can relate this to the $\epsilon$ coefficients via 
\[
C_{T,u}^{ee11} = 2 \left(\epsilon_{T,d}^{ee11} + \widetilde\epsilon_{T,d}^{ee11}\right)
 = 2\left(\epsilon_{T,d}^{ee11} + (\epsilon_{T,d}^{ee11})^* \right)
 = 4\,\mathrm{Re}\!\left(\epsilon_{T,d}^{ee11}\right),
\]
to conclude
\[
\xi^2_T/N^2F^2(Q^2) = 128\,\mathrm{Re}\!\left(\epsilon_{T,d}^{ee11}\right)^2
\left(\delta^n_d + \frac{Z}{N}\delta^p_d\right)^2.
\]
This enables us to translate the COHERENT bounds of \cite{AristizabalSierra:2018eqm}, $\xi_T/NF \leq 0.591$ at 90\%~CL, and $\xi_T/NF \leq 1.072$ at 99\%~CL to
\[
\begin{split}
\mathrm{Re}\!\left(\epsilon_{T,d}^{ee11}\right) &\leq 0.098 \quad \text{at 90\% CL,}\\
\mathrm{Re}\!\left(\epsilon_{T,d}^{ee11}\right) &\leq 0.178 \quad \text{at 99\% CL.}
\end{split}
\]
This compares to the current bound from beta decays \cite{Gonzalez-Alonso:2018omy}
\[\label{eq:tildeboundstensor}
|\widetilde\epsilon_{T,ud}^{ee11}|\leq 0.024\, \quad \text{at 90\% CL,}
\]
which is currently stronger by a factor of four. Projecting to near-future reactor-based Germanium experiments, $\xi_T/NF \leq 0.25$ \cite{Lindner:2016wff}, we find
\[
\mathrm{Re}\!\left(\epsilon_{T,d}^{ee11}\right) \leq 0.062 \quad \text{at 99\% CL,}
\]
which would be competitive with beta decays. Recall, however, that here we did not account for uncertainties in the tensor charges $\delta^n_q$ and $\delta^p_q$ which could weaken this projection, in particular in the case that $\delta^n_d$ is lower than the central value.

\begin{table}
{\centering
\begin{tabu}{X[$,l]X[$,l,2]X[$,l,2]X[c,2]}
\toprule
\text{Coefficient} & |C_X|\,[\Lambda = v/\sqrt{2}] & |C_X|\,[\Lambda = \SI{1}{TeV}] & Observable\\
\midrule
C_{ll}^{\mu eee} & \num{7.0e-7} &  \num{2.3e-5}\quad\text{\cite{Calibbi:2017uvl}} & $\mu\rightarrow eee$ \cite{sindrum} \\
C_{le}^{\mu eee, ee\mu e} & \num{1.0e-6} &\num{3.3e-5}\quad\text{\cite{Calibbi:2017uvl}} &  $\mu\rightarrow eee$ \cite{sindrum} \\
C_{ll}^{\tau eee} & \num{2.8e-4} & \num{9.2e-3}\quad\text{\cite{Calibbi:2017uvl}} & $\tau\rightarrow eee$ \cite{Hayasaka:2010np} \\
C_{le}^{\tau eee, ee\tau e} & \num{3.9e-4} & \num{1.3e-2}\quad\text{\cite{Calibbi:2017uvl}} & $\tau\rightarrow eee$ \cite{Hayasaka:2010np} \\
C_{ll}^{\tau \mu\mu\mu} & \num{2.4e-4} & \num{7.8e-3}\quad\text{\cite{Calibbi:2017uvl}} & $\tau\rightarrow \mu\mu\mu$ \cite{Hayasaka:2010np} \\
C_{le}^{\mu \tau\mu\mu,\mu\mu\mu\tau} & \num{3.3e-4} & \num{1.1e-2}\quad\text{\cite{Calibbi:2017uvl}} & $\tau\rightarrow \mu\mu\mu$ \cite{Hayasaka:2010np} \\
C_{le}^{\mu\mu\mu e, e\mu\mu\mu} & \num{5.5e-7} & \num{1.8e-4}\quad\text{\cite{Calibbi:2017uvl}} & $\mu\rightarrow e\gamma$ \cite{TheMEG:2016wtm} \\
C_{le}^{\mu\tau\tau e, e\tau\tau\mu} & \num{3.0e-7} & \num{1.0e-5}\quad\text{\cite{Calibbi:2017uvl}} & $\mu\rightarrow e\gamma$ \cite{TheMEG:2016wtm} \\
C_{lq(1)}^{e\mu 11} & \num{2.9e-8} & \num{9.4e-7}\quad\text{this work} & $\mu^{-}\mathrm{Au}\rightarrow e^{-}\mathrm{Au}$ \cite{sindrum2} \\
C_{lq(3)}^{e\mu 11} & \num{3.3e-7} & \num{1.1e-5}\quad\text{this work} & $\mu^{-}\mathrm{Au}\rightarrow e^{-}\mathrm{Au}$ \cite{sindrum2} \\
C_{lu}^{e\mu 11} & \num{6.0e-8} & \num{2.0e-6}\quad\text{\cite{Feruglio:2015gka}} & $\mu^{-}\mathrm{Au}\rightarrow e^{-}\mathrm{Au}$ \cite{sindrum2} \\
C_{ld}^{e\mu 11} & \num{5.2e-8} & \num{1.8e-6}\quad\text{\cite{Feruglio:2015gka}} & $\mu^{-}\mathrm{Au}\rightarrow e^{-}\mathrm{Au}$ \cite{sindrum2} \\
C_{elqd}^{e\mu 11} & \num{2.8e-8} & \num{9.2e-7}\quad\text{this work} & $\mu^{-}\mathrm{Au}\rightarrow e^{-}\mathrm{Au}$ \cite{sindrum2} \\
C_{eluq}^{e\mu 11} & \num{3.0e-8} & \num{9.7e-7}\quad\text{this work} & $\mu^{-}\mathrm{Au}\rightarrow e^{-}\mathrm{Au}$ \cite{sindrum2} \\
\midrule
C_{lNuq}^{ee 11} & \num{3.9e-4} &\num{1.3e-2}\quad \text{this work} & beta decay \cite{cirigliano2} \\
C_{lNqd}^{ee 11} & \num{4.0e-4} &\num{1.3e-2}\quad \text{this work} & beta decay \cite{cirigliano2} \\
C_{lNqd}^{'ee 11} & \num{2.4e-2} &\num{7.9e-1}\quad \text{this work} & beta decay \cite{Gonzalez-Alonso:2018omy} \\
\bottomrule
\end{tabu}
\caption{Summary low-energy single-parameter bounds (90\% CL) on SM(N)EFT operators used or obtained in this work. Bounds are given for GNI normalisation  $\Lambda=v/\sqrt{2}=(\sqrt{8}G_F)^{-1/2}$ and for $\Lambda=\SI{1}{TeV}$ as usual in SMEFT literature. The upper part corresponds to CLFV processes discussed in \autoref{sec:pheno-clfv}, while the lower part corresponds to first-generation exotic (pseudo)scalar and tensor interactions discussed in \autoref{sec:pheno-exotic}. 
}
\label{tab:eftbounds}
}
\end{table}

\section{Leptoquark models as UV completions}
\label{sec:leptoquarks}
Currently there is a renewed interest in \si{TeV}-scale leptoquark scenarios mainly inspired by the mild deviations from the SM expectation of $B$ meson decays, see e.g. 
\cite{Bauer:2015knc,Becirevic:2016yqi,Angelescu:2018tyl,Heeck:2018ntp}. They also appear in certain radiative neutrino mass models, e.g. \cite{Babu:2001ex,Deppisch:2016qqd,Chang:2016zll,Hati:2018fzc,Klein} and naturally arise in grand unified theories. Here, we are mainly discussing leptoquarks to demonstrate that the effective operators of \autoref{sec:operators} can be generated consistently in UV complete models. 

Including sterile neutrinos, the following general Lagrangian of fermions with possible scalar and vector leptoquarks can be obtained, which includes a few additional options with respect to the list in \cite{leptoquark1}. Like therein, we distinguish leptoquarks with fermion number $F=3B+L=0$ and $F=2$, where $B$ denotes baryon number and $L$ denotes lepton number:
\[
\begin{split}\label{eq:leptoquarks1}
\lagr_{F=2} &=
\left(s_{1L}\,\overline{q^c}i\tau_2l + s_{1e}\,\overline{u_R^c}e_R + s_{1N}\,\overline{d_R^c} N \right)S_1\\
&\quad + s_1'\,\overline{d_R^c}e\, S_1' + s_1''\,\overline{u_R^c}N \,S_1''
+s_{3}\overline{q^c}i\tau_2\vec\tau l\,\vec{S}_3
\\
&\quad +\left(v_{2R}\,\overline{q^c}^i\gamma_\mu e_R + v_{2L}
\,\overline{d_R^c}\gamma_\mu l^i\right)\epsilon_{ij}V_2^{\mu,j} \\
&\quad +\left(v_{2R}'\,\overline{q^c}^i\gamma_\mu N + v_{2L}'
\,\overline{u_R^c}\gamma_\mu l^i\right)\epsilon_{ij}{V_2^{\mu,j}}'
\plushc,
\end{split}
\]
\[
\begin{split}\label{eq:leptoquarks2}
\lagr_{F=0} &=
\left(r_{2R}\,\overline{q}^je_R + r_{2L}\,\overline{u_R}\,l^i\epsilon_{ij} \right)R_2^j\\
&\quad +\left(r_{2L}'\,\overline{d_R}\,l^i\epsilon_{ij} + r_{2R}'\,\overline{q}^j N \right){R_2^j}'\\
&\quad +\left(u_{1L}\,\overline{q}\gamma_\mu l + u_{1de}\,\overline{d_R}\gamma_\mu e_R + u_{1uN}\,\overline{u_R}\gamma_\mu N\right)U_1^\mu\\
&\quad +u_1'\,\overline{u_R}\gamma_\mu e_R\, {U_1^\mu}'
+u_1''\,\overline{d_R}\gamma_\mu N\, {U_1^\mu}''
+u_{3}\,\overline{q}\,\vec\tau \gamma_\mu\,l\,\vec{U}_3^\mu
\plushc
\end{split}
\]
With respect to \cite{leptoquark1}, the identification reads
\[
S_1,S_1',S_3,V_2,V_2',R_2,R_2',U_1,U_1',U_3 \longleftrightarrow
S_1,\widetilde{S}_1,S_3,V_2,\widetilde{V}_2,R_2,\widetilde{R}_2,U_1,\widetilde{U}_1,U_3\,,
\]
while $S_1''$ and $U_1''$ are new. We list their quantum numbers in \autoref{tab:leptoquarks}. In general (not assuming baryon number conservation), there are additional couplings to be considered for some leptoquarks:
\[\label{eq:leptoquarks3}
\begin{split}
\lagr_{F=2}^{\Delta B}&=
s_{1B}\overline{q}i\tau_2q^c S_1 + 
s_{1B}'\overline{u}u^c S_1' +
s_{1B}''\overline{d}d^c S_1'' \\
&\quad +
s_{3B}\overline{q}\vec\tau i\tau_2q^c \vec{S}_3 
 +\overline{q}\gamma_\mu u^c V_2^\mu 
+\overline{q}\gamma_\mu d^c {V_2^\mu}',
\end{split}
\]
which, together with \eqref{eq:leptoquarks1} may lead to proton decay, such that to avoid strong bounds usually baryon number conservation is assumed.\footnote{Of course, if one took only the couplings of \eqref{eq:leptoquarks3} and not of \eqref{eq:leptoquarks1}, one could assign definite baryon numbers to the new fields and recover baryon number conservation. Since then there are no couplings to leptons, however, we are not interested in this case.}
Assuming large masses, those leptoquarks can potentially lead to the operators in \autoref{tab:4fermionSMEFT} after integrating them out. We will not discuss all details, but list which leptoquarks can lead to which operators, assuming only one leptoquark type being present at a time, in the rightmost column of \autoref{tab:leptoquarks}.

\begin{table}
{\centering
\begin{tabu}{X[$,l,0.5]X[$,c]X[$,c]X[$,c]X[$,c]X[$,c]X[4,$,l]}
\toprule
 & F &\text{Spin} & \mathrm{SU(3)}_\mathrm{C} & \mathrm{SU(2)}_L & \mathrm{U(1)}_Y &\text{Dimension-6 operators}\\
\midrule
S_1 	&-2 &0 &\overline{3} & 1 &2/3 & \mathcal{O}_{lq}^{(1)}, \mathcal{O}_{Nd},\mathcal{O}_{lNqd},\mathcal{O}_{lNqd}',\\
&&&&&&\mathcal{O}_{eluq},\mathcal{O}_{eNud}\\
S_1'  &-2 &0 &\overline{3} & 1 &8/3\\
S_1''  &-2 &0 &\overline{3} & 1 &-4/3 & \mathcal{O}_{Nu}\\
S_3  &-2 &0 &\overline{3} & 3 &2/3 & \mathcal{O}_{lq}^{(3)}\\
V_2  &-2 &1 &\overline{3} & 2 &5/3 & \mathcal{O}_{ld},\mathcal{O}_{elqd}\\
V_2' &-2 &1 &\overline{3} & 2 & -1/3 & \mathcal{O}_{Nq},\mathcal{O}_{lu},\mathcal{O}_{lNuq}\\
R_2  &0 &0 &3 & 2 &7/3 & \mathcal{O}_{lu},\mathcal{O}_{eluq}\\
R_2' &0 &0 &3 & 2 &1/3 & \mathcal{O}_{ld},\mathcal{O}_{Nq},\mathcal{O}_{lNqd},\mathcal{O}_{lNqd}'\\
U_1  &0 &1 &3 & 1 &4/3 & \mathcal{O}_{lq}^{(1)},\mathcal{O}_{Nu},\mathcal{O}_{elqd},\mathcal{O}_{lNuq},\mathcal{O}_{eNud}\\
U_1'  &0 &1 &3 & 1 &10/3\\
U_1''  &0 &1 &3 & 1 &-2/3 & \mathcal{O}_{Nd}\\
U_3  &0 &1 &3 & 3 &4/3 & \mathcal{O}_{lq}^{(3)}\\
\bottomrule
\end{tabu}
\caption{Coupling constants and operators appearing in generic neutral-current \eqref{eq:epsilonNC} and charged-current Lagrangians \eqref{eq:epsilonCC}.}
\label{tab:leptoquarks}
}
\end{table}

We would like to draw attention to a particular combination of two leptoquarks. It has been found that, on the one hand, introducing $S_1$ (or $S_3$) and $R_2'$ can explain $B$ physics anomalies, and on the other hand radiative neutrino masses could be generated at one-loop level\cite{Cata:2019wbu}, where the couplings of \eqref{eq:leptoquarks3} are avoided by assuming baryon number conservation. This is possible because the particular quantum numbers allow for a coupling term
\[\label{eq:higgscoupling}
\lagr=\mu\,S_1\phi^\dagger R_2'\,,
\]
and similarly for $S_3$. In fact this is the only combination of two scalar (color-triplet) leptoquarks that can generate radiative neutrino masses without a second copy that appears in a 3-leptoquark coupling \cite{Klein}.\footnote{If one admits sextet leptoquarks, there is the further possibility of combining $S_1$ with a (6,1,4/6) representation as discussed in \cite{Chang:2016zll}.} 
Considering \eqref{eq:leptoquarks1} and \eqref{eq:leptoquarks2}, the only other combination which admits a one-loop neutrino mass via a coupling like \eqref{eq:higgscoupling} is constituted by the two vectors $U_1$ and $V_2'$, as discussed in \cite{Deppisch:2016qqd}. 

Now it is an interesting observation that it is exclusively $S_1$ and $R_2'$ that potentially produce the operators $\mathcal{O}_{lNqd}$ and $\mathcal{O}_{lNqd}'$ responsible for $\widetilde\epsilon_{S/P/T,ud}$ and $\epsilon_{S/P/T,d}$. Meanwhile the other combination, $U_1$ and $V_2'$ are the only ones giving rise to the operator $\mathcal{O}_{lNuq}$ responsible for $\widetilde\epsilon_{S/P,ud}$ and $\epsilon_{S/P,u}$. Namely, it is precisely the leptoquarks that are able to produce radiative neutrino masses which can produce \emph{those} semileptonic (pseudo)scalar and tensor operators that are safe from constraints from charged-lepton interactions (the operators in \autoref{tab:coefficientsNC} and \autoref{tab:coefficientsCC} not printed in red). 

Let us expand on this further by trying to recover the GNI scenarios of \autoref{sec:pheno-exotic}. 
Integrating out an $S_1$ leptoquark and Fierz transforming the resulting expression produces the operators $\mathcal{O}_{lNqd}$ and $\mathcal{O}_{lNqd}'$ with
\[
\label{eq:leptoquark-gni}
C_{lNqd}^{ee11} = 4C_{lNqd}^{'ee11} \sim \frac{\left(s_{1L}^{1e}\right)^*s_{1N}^{1e}}{\sqrt{8}G_F\, m_{S_1}^2}\,
\sim 10^{-4} s_{1N}^{1e}\,,
\]
where we use $s_{1L}^*\sim 10^{-3}$ and $m_{S_1}^2\sim\SI{1}{TeV}$ motivated by the radiative neutrino mass scenario in \cite{Cata:2019wbu}. 
This corresponds to a scenario similar to the one described in \autoref{sec:pheno-singlescalar}, just with $C_{lNqd}$ taking the role of $C_{lNuq}$, combined with the scenario of \autoref{sec:pheno-tensor}. Applying the direct bounds \eqref{eq:tildebounds} and \eqref{eq:tildeboundstensor} from beta decay on $\widetilde\epsilon_{P,ud}^{ee11}$ and $\widetilde\epsilon_{T,ud}^{ee11}$ we can cast the experimental bounds\footnote{Strictly speaking those bounds are obtained assuming only one GNI parameter at a time, but we assume here that their measurements are sufficiently independent.}
\[
\begin{split}
|C_{lNqd}^{ee11}| &\leq \num{4.0e-4}\,,\\
|C_{lNqd}^{'ee11}| &\leq \num{2.4e-2}\,.
\end{split}
\] 
Hence, with an order one or smaller coupling $s_{1,N}$ of sterile neutrinos to $S_1$, this scenario can plausibly generate pseudoscalar neutrino interactions which are compatible with current beta decay bounds but close enough to be potentially detectable in the near future. On the other hand, at the same time the tensor interactions would then also be of order $10^{-4}$ and thus two orders of magnitude below the current limit \eqref{eq:tildeboundstensor}.

The setup described in \autoref{sec:pheno-twoscalars} with both $C_{lNqd}$ and $C_{lNuq}$ being larger order than $10^{-4}$ and cancelling in $\widetilde\epsilon_P^{ee11}$ is very unlikely to be realised together with the radiative neutrino masses due to \eqref{eq:leptoquark-gni}.
However, if we abandon the requirement that radiative neutrino masses are produced by the same leptoquark as the GNI, the two-operator scenario of  \autoref{sec:pheno-twoscalars} can in principle be realised with suitable couplings of e.g.\ $S_1$ and $V_2'$.

\section{Conclusions}
\label{sec:conclusions}

In this article, we investigated General Neutrino Interactions as parametrised in Eqs.\ (\ref{eq:epsilonNC}, \ref{eq:epsilonCC}), which describe all possible Lorentz-invariant neutral- and charged-current interactions with other SM fermions. 
We discussed the properties of those  interactions under the assumption that they originate from gauge invariant dimension-six operators composed of four SM fermion representations and (light) sterile neutrinos. Those operators  parametrise heavy new physics above the weak scale. We found that almost all General Neutrino Interaction parameters can be generated in this way, in contrast to SMEFT without sterile neutrinos. Considering  phenomenological aspects, we demonstrated that, in this framework, constraints on CLFV forbid sizable neutrino flavour changing vector-like interactions, while some chirality flipping scalar- and tensor-like interactions are still viable. On the other hand, we showed that there can be interesting and potentially detectable correlations between scalar and tensor interactions of neutrinos with the first generation of quarks. In particular, the same operators can generate neutral-current interactions detectable in CE$\nu$NS, and charged-current interactions detectable in beta decays. Finally, we discussed that the operators responsible for GNI can originate in UV-complete models from heavy leptoquarks. 

We conclude that viable scenarios implying new neutrino interactions of scalar and tensor type detectable at low-energy experiments exist. At the same time vector-like NSI are generally constrained below near-future detector reach in  SMEFT scenarios. In phenomenological analyses of low-energy data, all GNI should be included.

\subsection*{Acknowledgements}
IB is supported by the IMPRS-PTFS. WR is supported by the DFG with grant RO 2516/7-1 in the Heisenberg program.

\appendix

\section{General neutral-current interactions}
\label{sec:details}
In this section, for convenience and future reference, we list, for both common parametrisations of general neutral-current neutrino interactions, the parameters and numbers of degrees of freedom in general, in the case of Majorana neutrinos, and in the case of CP symmetry.
As the \emph{epsilon parametrisation} we refer to
\[\label{eq:lagrangian}
\lagr=-\frac{G_F}{\sqrt{2}}\sum_{\alpha,\beta,\gamma,\delta}\sum_{j=1}^{10}\left(\stackrel{\phantom{j}(\sim)j}{\epsilon}\right)_{\abgd}
\left(\overline{\nu}_\alpha\mathcal{O}_j\nu_\beta\right)
\left(\overline{f}_\gamma\mathcal{O}_j'f_\delta\right),
\]
with operators explained in \autoref{tab:chiral-operators}. As the \emph{CD parametrisation} we refer to the equivalent formulation
\begin{align}\label{eq:CDlagrangian}
\lagr&=-\frac{G_F}{\sqrt{2}}\sum_{a=S,P,V,A,T}\left(\overline{\nu}_\alpha\,\Gamma^a\nu_\beta\right)
\left(\overline{f_\gamma}\Gamma^a(C^a_{\abgd}+\overline{D}^a_{\abgd}i\gamma^5)f_\delta\right),
\end{align}
where the five possible independent combinations of Dirac matrices are defined as
\[
\Gamma^a\in \left\{\mathds{1},i\gamma^5,\gamma^\mu,\gamma^\mu\gamma^5,\sigma^{\mu\nu}\right\},
\]
and the ten coefficients $C^a_{\abgd}$ and
\[\label{eq:CaDa}
D^a_{\alpha\beta\gamma\delta}\equiv\begin{cases}
\overline{D}^a_{\abgd} & (a=S,P,T) \\
i\overline{D}^a_{\abgd}& (a=V,A)\end{cases}
\]
obey the relations stated below. Note that
\[
\gamma^0(\Gamma^a)^\dagger\gamma^0 = \Gamma^a\,,
\]
such that 
\[
\left(\overline\psi_{\alpha} \Gamma^a \psi_\beta\right)^\dagger = \overline\psi_{\beta} \Gamma^a \psi_\alpha\,
\]
for generic Dirac spinors $\psi_\alpha,\psi_\beta$.
The two parametrisations are related by
\[\label{eq:epsilontocd}
\begin{split}
\epsilon^L &= \frac14\left(C^V-D^V+C^A-D^A\right),\\
\epsilon^R &= \frac14\left(C^V+D^V-C^A-D^A\right),\\
\epsilon^S&=\frac12\left(C^S+iD^P\right),\\
-\epsilon^P&=\frac12\left(C^P+iD^S\right),\\
\epsilon^T&=\frac14\left(C^T-iD^T\right),
\end{split}
\qquad
\begin{split}
\widetilde\epsilon^L &= \frac14\left(C^V-D^V-C^A+D^A\right),\\
\widetilde\epsilon^R &= \frac14\left(C^V+D^V+C^A+D^A\right),\\
\widetilde\epsilon^S&=\frac12\left(C^S-iD^P\right),\\
-\widetilde\epsilon^P&=\frac12\left(-C^P+iD^S\right),\\
\widetilde\epsilon^T&=\frac14\left(C^T+iD^T\right),
\end{split}
\]
where flavour indices are suppressed. Note that the relation between the two parametrisations is the same for charged-current interactions.

The following properties of the coefficients under particular assumptions can be straightforwardly derived.
\begin{enumerate}
\item \textbf{In general:} (810 real parameters)
\[
\begin{split}
\epsilon^j_{\abgd}&=\epsilon^{j*}_{\badg}\,,\quad j=L,R,\widetilde{L},\widetilde R\,,\\
\epsilon^{S}_{\abgd}&=\widetilde\epsilon^{S*}_{\badg}\,,\\
\epsilon^{P}_{\abgd}&=-\widetilde\epsilon^{P*}_{\badg}\,,\\
\epsilon^{T}_{\abgd}&=\widetilde\epsilon^{T*}_{\badg}\,,\\
\end{split}
\qquad\qquad
\begin{split}
C^a_{\abgd} &= C^{a*}_{\badg}\,,\\
\quad D^a_{\abgd} &= D^{a*}_{\badg}\,.
\end{split}
\]

\item \textbf{CP invariance:} (423 real parameters)\\
\[
\begin{split}
\epsilon^j_{\abgd}&\in\mathbb{R}\qquad\qquad \forall j,\\
\epsilon^j_{\abgd}&=\epsilon^j_{\badg}\qquad j=L,R,\widetilde{L},\widetilde R\,,
\end{split}
\qquad\qquad
\begin{split}
C^a_{\abgd}&\in\mathbb{R}\,\qquad\quad \forall a,\\
D^a_{\abgd}&\in\mathbb{R}\,,\qquad \phantom{i}a=V,A\,,\\
D^a_{\abgd}&\in i\mathbb{R}\,,\qquad a=S,P,T\,.
\end{split}
\]

\item \textbf{Majorana neutrinos:} (432 real parameters)\\
\[
\begin{split}
\epsilon^j_{\abgd} &= - \widetilde\epsilon^j_{\beta\alpha\gamma\delta} \,,\qquad j=L,R\\
\epsilon^k_{\abgd} &= \epsilon^k_{\beta\alpha\gamma\delta}
\,,\qquad \phantom{-}k=S,P\\
\epsilon^T_{\abgd} &= -\epsilon^T_{\beta\alpha\gamma\delta}\,.
\end{split}
\qquad\qquad
\begin{split}
C^a_{\abgd}&=C^a_{\beta\alpha\gamma\delta}\,,\qquad
\,\\
\phantom{-}D^a_{\abgd}&=D^a_{\beta\alpha\gamma\delta}\,,\qquad
a=S,P,A\,,\\
C^b_{\abgd}&=-C^b_{\beta\alpha\gamma\delta}\,,\qquad
\\
D^b_{\abgd}&=-D^b_{\beta\alpha\gamma\delta}\,,\qquad
b=V,T\,.
\end{split}
\]

\section{Details on the Higgs-fermion operators}
\label{sec:mixedoperators}
In this section, we summarise the effects of the mixed operators in the third column of \autoref{tab:bosonSMEFT}. If the operators are evaluated at the Higgs vacuum expectation value, one finds the following modified weak interaction Lagrangians,
\begin{align}
\lagr_Z &= -\frac{g}{2c_W}Z_\mu j^\mu_Z\,, \label{eq:nc}\\
\lagr_W &= -\frac{g}{2\sqrt{2}}W_\mu j^\mu_W\plushc, \label{eq:cc}
\end{align}
where $g$ denotes the SU(2)$_L$ gauge coupling and $c_W$ the cosine of the weak mixing angle.
The modified currents read
\begin{align}
\nonumber
j^\mu_Z &= \left(\delta^{\alpha\beta}-2C^{\alpha\beta}_{\varphi l(1)}+2C^{\alpha\beta}_{\varphi l(3)}\right)\overline{\nu}_\alpha\gamma^\mu \nu_\beta -2C^{\alpha\beta}_{\varphi N}\overline{N}_\alpha\gamma^\mu N_\beta
\\
&\quad+\left((-1+2s_W^2)\delta^{\alpha\beta}-2C^{\alpha\beta}_{\varphi l(1)}-2C^{\alpha\beta}_{\varphi l(3)}\right)\overline{e}_\alpha\gamma^\mu P_L e_\beta \\
&\qquad 
+ 2s_W^2\delta^{\alpha\beta}\overline{e}_\alpha\gamma^\mu P_R e_\beta
+j^\mu_{Z,q}\,,\nonumber
\\
j_W^\mu &= (2\delta^{\alpha\beta}+4C^{\alpha\beta}_{\varphi l(3)})\overline{\nu}_\alpha\gamma^\mu P_Le_\beta+2C^{\alpha\beta}_{\varphi Ne}\overline{N}_\alpha\gamma^\mu P_Re_\beta
+ j_{W,q}^\mu\,,
\end{align}
and we included all SM contributions, $j_{Z,q}$ and $j_{W,q}$ denoting the SM quark currents. Note that two distinctively new features arise, namely a right-handed charged-current coupling to $W$, and a right-handed neutrino neutral-current coupling to $Z$. 

These modified couplings result in the following low-energy four-fermion interaction parameters (at first order in operator coefficients).
It is useful to employ the SM gauge boson couplings
\[
\begin{split}
g_L^e &= -\frac12 + s_W^2\,,\\
g_R^e &= s_W^2\,,\phantom{\frac12}
\end{split}
\qquad
\begin{split}
g_L^u &= \frac12-\frac23s_W^2\,,\\
g_R^u &= -\frac23 s_W^2\,,
\end{split}
\qquad
\begin{split}
g_L^d &= -\frac12 + \frac13 s_W^2\,,\\
g_R^d &= \frac13 s_W^2\,.
\end{split}
\]
Due to possible Fierz transformations, the left-handed electron coupling has several contributions,
\[
\begin{split}
\epsilon_{L,e}^{\abgd}&=2\delta^{\alpha\delta}C^{\gamma\beta}_{\varphi l(3)} + 2C^{\alpha\delta}_{\varphi l(3)}\delta^{\gamma\beta}
\\
&\quad -\delta^{\alpha\beta}\left(C^{\gamma\delta}_{\varphi l (1)}+C^{\gamma\delta}_{\varphi l (3)}\right)
-2g_L^e\delta^{\gamma\delta}\left(C^{\alpha\beta}_{\varphi l (1)}-C^{\alpha\beta}_{\varphi l (3)}\right).
\end{split}
\]
The left-handed quark neutral-current coupling and the right-handed neutral-current couplings to $f=u,d,e$ arise simply (and all simultaneously) from the modified $Z$ coupling of neutrinos, 
\[
\begin{split}
\epsilon_{L,q}^{\abgd} &=
2g_L^q\delta^{\gamma\delta}\left(C^{\alpha\beta}_{\varphi l (3)}-C^{\alpha\beta}_{\varphi l (1)}\right), \\
\epsilon_{R,f}^{\abgd} &=
2g_R^f\delta^{\gamma\delta}\left(C^{\alpha\beta}_{\varphi l (3)}-C^{\alpha\beta}_{\varphi l (1)}\right), \\
\end{split}
\]
where $q=u,d$.
Also simultaneously the modified $Z$ coupling induced by $\mathcal{O}_{\varphi N}$ results in
\[
\begin{split}
\widetilde\epsilon_{L,f}^{\abgd} &=
-2g_L^f\delta^{\gamma\delta}C^{\alpha\beta}_{\varphi N}\,,\\
\widetilde\epsilon_{R,f}^{\abgd} &=
-2g_R^f\delta^{\gamma\delta}C^{\alpha\beta}_{\varphi N}\,.
\end{split}
\]
The right-handed $W$ coupling induced by $\mathcal{O}_{\varphi Ne}$ implies
\[
\begin{split}
\epsilon_{S,e}^{\abgd}&=\epsilon_{P,e}^{\abgd}
=2\delta^{\beta\gamma}C^{\alpha\delta}_{\varphi Ne}\,, \\
\widetilde\epsilon_{L,ud}^{\abgd}
&= (C^{\beta\alpha}_{\varphi Ne})^*\delta^{\gamma\delta},
\end{split}
\]
and finally another semileptonic charged-current term arises,
\[
\epsilon_{L,ud}^{\abgd}
=2\delta^{\gamma\delta}(C^{\beta\alpha}_{\varphi l(3)})^*.
\]
We note that besides the neutrino interactions, the left-handed electron coupling to $Z$ is affected by two operators, namely
\[
\begin{split}
\lagr_{Ze} = -\frac{g}{2c_W}Z_\mu {j^\mu_{Ze}}'
&=-\frac{g}{2c_W}Z_\mu 
\left(2g_L^e\delta^{\alpha\beta}-2C^{\alpha\beta}_{\varphi l(1)}-2C^{\alpha\beta}_{\varphi l(3)}\right)\overline{e}_\alpha\gamma^\mu P_L e_\beta \\
&\equiv -\frac{g}{2c_W}Z_\mu\,(2{g_L^e}')^{\alpha\beta}\overline{e}_\alpha\gamma^\mu P_L e_\beta\,,
\end{split}
\]
i.e.\ $g_L^e$ gets replaced by a flavour dependent coupling $({g_L^e}')^{\alpha\beta}$ in \emph{all} neutral-current fermionic interactions. Note also, however, that a cancellation of Wilson coefficients may suppress this modified coupling.

\end{enumerate}

\bibliographystyle{utphys}
\bibliography{bibliography.bib}{}
\end{document}